\documentclass[galaxies,review,accept,pdftex,moreauthors]{Definitions/mdpi} 

\firstpage{1} 
\pubvolume{1}
\issuenum{1}
\articlenumber{0}
\pubyear{2025}
\copyrightyear{2025}
\externaleditor{Margo Aller}
\datereceived{14 March 2025} 
\daterevised{14 April 2025} 
\dateaccepted{16 April 2025} 
\datepublished{30 April 2025 } 
\hreflink{https://doi.org/} 

\usepackage{changepage}
\usepackage{comment}
\usepackage{fnpct}
\usepackage[normalem]{ulem}
\usepackage{gensymb}

\newcommand{\ion}[2]{#1~{\footnotesize \textsc{#2}}}
\newcommand{\msun}{\mathrm{M_\odot}}
\newcommand{\ud}{\mathrm{d}}
\newcommand{\tn}{T_{90}}
\newcommand{\mnras}{Mon.~Not. R. Astron.~Soc.}
\newcommand{\apj}{Astrophys. J.}
\newcommand{\aap}{Astron. Astrophys.}
\newcommand{\aj}{Astron. J.}
\newcommand{\actaa}{Acta. Astron.}
\newcommand{\pasp}{Publ. Astron. Soc. Pac.}

\newcommand{\apjs}{Astrophys. J. Suppl.}
\newcommand{\araa}{Annu. Rev. Astron. Astrophys.}

\newcommand{\apjl}{Astrophys. J. Lett.}
\newcommand{\nat}{Nature}
\newcommand{\prd}{Phys. Rev. D}

\newcommand{\physrep}{Phys.~Rep.}
\newcommand{\cjaa}{Chin. J. Astron. Astrophys.}
\newcommand{\ssr}{Space Sci. Rev.}

\Title{Exploring the GRB--Supernova Connection
	: Does a Superluminous Hypernova Population Exist?}

\TitleCitation{Exploring the GRB--Supernova Connection: Does a Superluminous Hypernova Population Exist?}

 \Author{Achille Fiore $^{1,2,3}$*\orcidA{}, Ludovica Crosato Menegazzi~$^{4}$\orcidC{} and Giulia Stratta~$^{2,5,6}$\orcidB{}}
\AuthorNames{Achille Fiore*, Ludovica Crosato Menegazzi, Giulia Stratta}

\AuthorCitation{Fiore, A.; Crosato Menegazzi, L.; Stratta, G.}

\address{%
$^{1}$ \quad INAF-Osservatorio Astronomico d’Abruzzo, Via Mentore Maggini Snc, 64100 Teramo, Italy\\
$^{2}$ \quad Institut f\"ur Theoretische Physik, Goethe Universit\"at, Max-von-Laue-Str. 1, 60438 Frankfurt am Main, Germany;  giulia.stratta@inaf.it\\
$^{3}$ \quad INAF-Osservatorio Astronomico di Padova, Vicolo dell’Osservatorio 5, 35122 Padova, Italy\\
$^{4}$ \quad Max Planck Institute for Gravitational Physics (Albert Einstein Institute), Am M{\"u}hlenberg 1, \mbox{Potsdam 14476, Germany; ludovica.crosatomenegazzi@aei.mpg.de} \\
$^{5}$ \quad Istituto di Astrofisica e Planetologia Spaziali, via Fosso del Cavaliere 100, 00133 Roma, Italy\\
$^{6}$ \quad INAF-Osservatorio Astronomico di Bologna, via P. Gobetti 93/3, 40129 Bologna, Italy\\
}
\corres{Correspondence: achille.fiore@inaf.it}

\abstract{Observations of several gamma-ray bursts (GRBs) that are temporally and spatially compatible with energetic supernovae (hypernovae) have established their common origin. In one case (GRB~111209A/SN~2011kl), the associated supernova was classified as superluminous (SN~2011kl). The exceptional duration of the observed gamma-ray prompt emission of GRB~111209A (about 7 h) is widely considered key to unlocking the physics behind the still mysterious origin of superluminous supernovae (SLSNe). 
We review the main observational and theoretical findings that may link some ultra-long GRBs to SLSNe. Specifically, we examine notable events and the role of progenitors and host galaxies in shaping these phenomena and focus on the proposed models. While a magnetar central engine is a plausible mechanism for both luminous and long-duration GRBs, a conclusive answer remains elusive, as alternative explanations are still viable. Further observational and theoretical work is required to clarify  progenitor pathways and explosion mechanisms, potentially extending the classical GRB-SN connection to rare superluminous hypernovae.}


\begin{document}

\section{Introduction}
\begin{figure}[H]
    
    \includegraphics[width=0.9\textwidth]{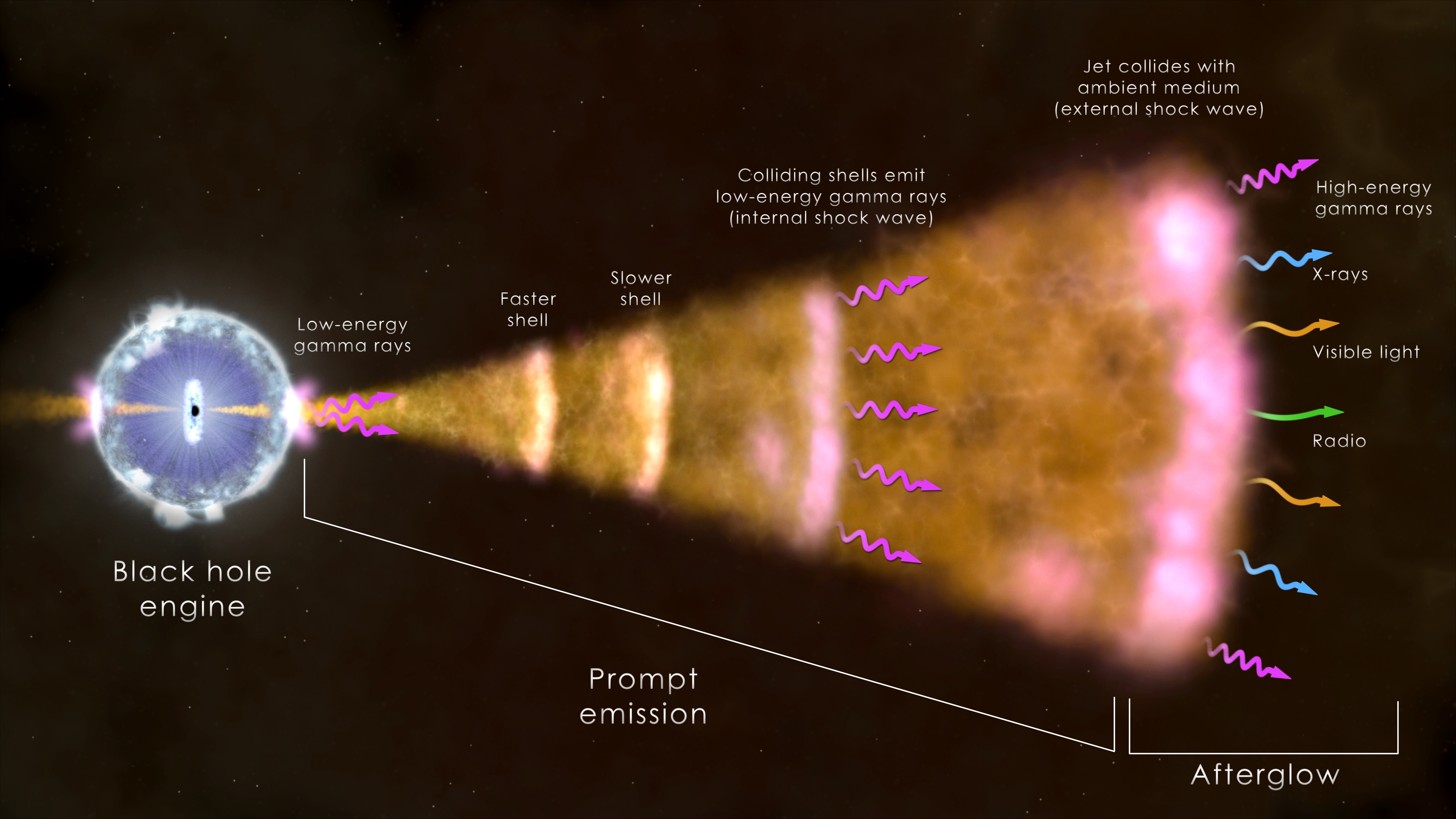}
    \caption{A
 schematic representation of a GRB illustrating the physical mechanisms producing the prompt and the afterglow emissions due to internal and external shocks, respectively. Credit: NASA's Goddard Space Flight Center.}
    \label{fig:Nasa_Goddard}
\end{figure}

GRBs are among the most luminous and energetic phenomena observable in the Universe. They arise as a prompt emission in the keV-MeV energy bands, followed by an afterglow emission, the latter of which spreads its power over longer wavelengths (see Figure~\ref{fig:Nasa_Goddard} for a schematic representation of GRB components). The prompt emission usually lasts from a few seconds up to minutes and is bimodally distributed, identifying two classes of GRBs---long GRBs (LGRBs) and short GRBs (SGRBs)~\cite{kouveliotouetal1993}---depending on whether their duration is shorter or longer than 2 s (see also Section~\ref{sec:grbs}). 

In most cases, the 2 s boundary discerns two different astrophysical phenomena. In fact, SGRBs have been found to be counterparts of compact binary mergers, while LGRBs are thought to occur with the death of massive stars. This review focuses on LGRBs and the properties of their associated SNe---in particular, on the putative extension of this general paradigm to the longest-lived and brightest events.

As is well known from stellar evolution theory, massive stars ($M_{\rm ZAMS}\gtrsim8\,\msun$~\cite{smartt2009}) undergo the collapse of their degenerate cores and can be progenitors of core-collapse SNe; in a number of cases, LGRBs have been observed to occur in spatial and temporal coincidence with core-collapse SNe (CCSNe). The very first documented case of the GRB--SN connection dates back to 1998 and it was reported by Galama~ et~al.~\cite{galamaetal1998}, who observed the energetic and luminous LGRB GRB 980425 associated with SN~1998bw. In 1998, Bohdan Paczy\'nski reported on the very bright afterglow of GRB980425, which largely outshone SN~1998bw, as well as every known SN type:

\begin{quote}
`Therefore, it seems appropriate to call it a hypernova.'~\cite{paczynski1998}
\end{quote}
\begin{flushright}
     \footnotesize{© AAS. Reproduced with permission.}
\end{flushright}

\textls[-15]{
At that time, the term ``hypernova'' was introduced to describe the whole GRB+afterglow event. Since then, this term has been been used to refer to another longer-wavelength counterpart of the GRB emission, i.e., a very energetic {``}broad-lined{''} type-Ic SN (SN Ic BL, see Section~\ref{sec:snzoo})
, whose spectroscopic features are reminiscent of a stripped progenitor like SNe Ic but whose line broadening reveals high expansion velocities  (see Section~\ref{sec:snzoo}). }

At the time of writing, about $45$ LGRBs have been found in association with SNe~\cite{belkinandpozanenko2023}. Interestingly, there are GRB~SNe that are not classified as SNe Ic BL and, in one case, much brighter than ordinary hypernovae. In fact, LGRB GRB 111209A was associated with SN~2011kl~\cite{greineretal2015}, classified as hydrogen-poor SLSNe (see Section~\ref{sec:slsne}). The association between GRB 111209A and SN~2011kl has become an archetype for a postulated LGRB--SLSN connection. In the following, we review the basic properties of LGRBs and SLSNe, as well as the chief observational and theoretical arguments supporting and/or disfavoring their affinity. The work is structured as follows: In Section~\ref{sec:protagonists}, we provide an overview of LGRBs and SLSNe from observational and theoretical perspectives. In Section~\ref{sec:magnetars} we review the properties of magnetars. In Section~\ref{sec:both}, we briefly describe how the magnetar scenario can explain both (SL)SNe and LGRBs, eventually in a single event. In Section~\ref{sec:pecgrbsne}, we examine some notable cases of peculiar GRB--SN associations eventually involving luminous SNe and/or very long-lasting LGRBs. In Section~\ref{sec:environments}, the role of their environment is summarized. We collect our final remarks in Section~\ref{sec:discussion}.

In the following, physical quantities are generally scaled to reference values using the conventional subscript notation. For instance, a generic quantity ($Q$) can be denoted as $Q_x$, which is defined as $Q/x$ (or $Q/10^x$ in the case of quantities that are expressed as~exponents).
\subsection{An Interlude: $\tn$ as a Proxy for the GRB Duration}
\label{sec:duration}
Historically, the duration of the high-energy prompt emission of a GRB has been identified as a first indication of the GRB progenitor's nature. A widely used proxy for it is the so-called $\tn$, corresponding to the time interval in which the fraction of the total counts collected during the high-energy burst grows from 5\% to 95\%. There is also a clear distinction in the spectral hardness, as LGRB spectra are usually softer than those of short GRBs
. The distribution of $\tn$ over large samples of GRBs clearly distinguishes between the shortest and longest GRBs (see e.g., Figure~4 in~\cite{minaevandpozanenko2020}).
 The empirical short/long GRB dichotomy 
 is theoretically motivated, as SGRBs are expected from NS mergers, while LGRBs are more naturally explained in the core collapse of massive stars (see Introduction). However, the use of $\tn$ as a discriminator (SGRBs and LGRBs are those with $\tn<2$ s and $\tn>2$ s, respectively~\cite{kouveliotouetal1993}) might be questionable due to the dependence of the $\tn$ on the energy band in which it is measured
 \endnote{Typically, a $\tn$ measured in softer bands is longer.}, as well as other factors, e.g., the choice of the $\gamma$-ray sky background model\endnote{The background level has to be modeled (usually with a polynomial function) to be subtracted from
 	 the counts, and different choices could potentially affect $\tn$ determination~\cite{minaevandpozanenko2010}.}
 and the sensitivity of the detector. 
In general, besides the GRB duration and spectral properties, other indicators, such as the host galaxy's age, the site of the GRB within the galaxy, the presence or absence of an associated kilonova or SN, are relevant to the identification of the progenitor. For this reason, rather than ``short'' and ``long'', GRBs are often classified as ``Type I'' and ``Type II'', corresponding to merger- and collapsar-driven GRBs, respectively \cite[]{Zhang2009ApJ...703.1696Z}.
Additional progenitor indicators have been proposed on the basis of the general spectral and temporal properties of the prompt emission. For instance, one was discussed by Minaev and Pozanenko~\cite{minaevandpozanenko2020} based on the Amati relation~\cite{amatietal2002,amati2006,amatietal2008,amatietal2009}, according to which Type I and Type II obey two different $E_{\rm iso}-E_{\rm p}$ relations (see {Figure~4 in~\cite{minaevandpozanenko2020}}), where $E_{\rm iso}$ is the total isotropic equivalent energy emitted by the GRB and $E_{\rm p}$ is its rest-frame peak energy. 
 
\begin{figure}[H]
    
    \includegraphics[width=0.85\textwidth]{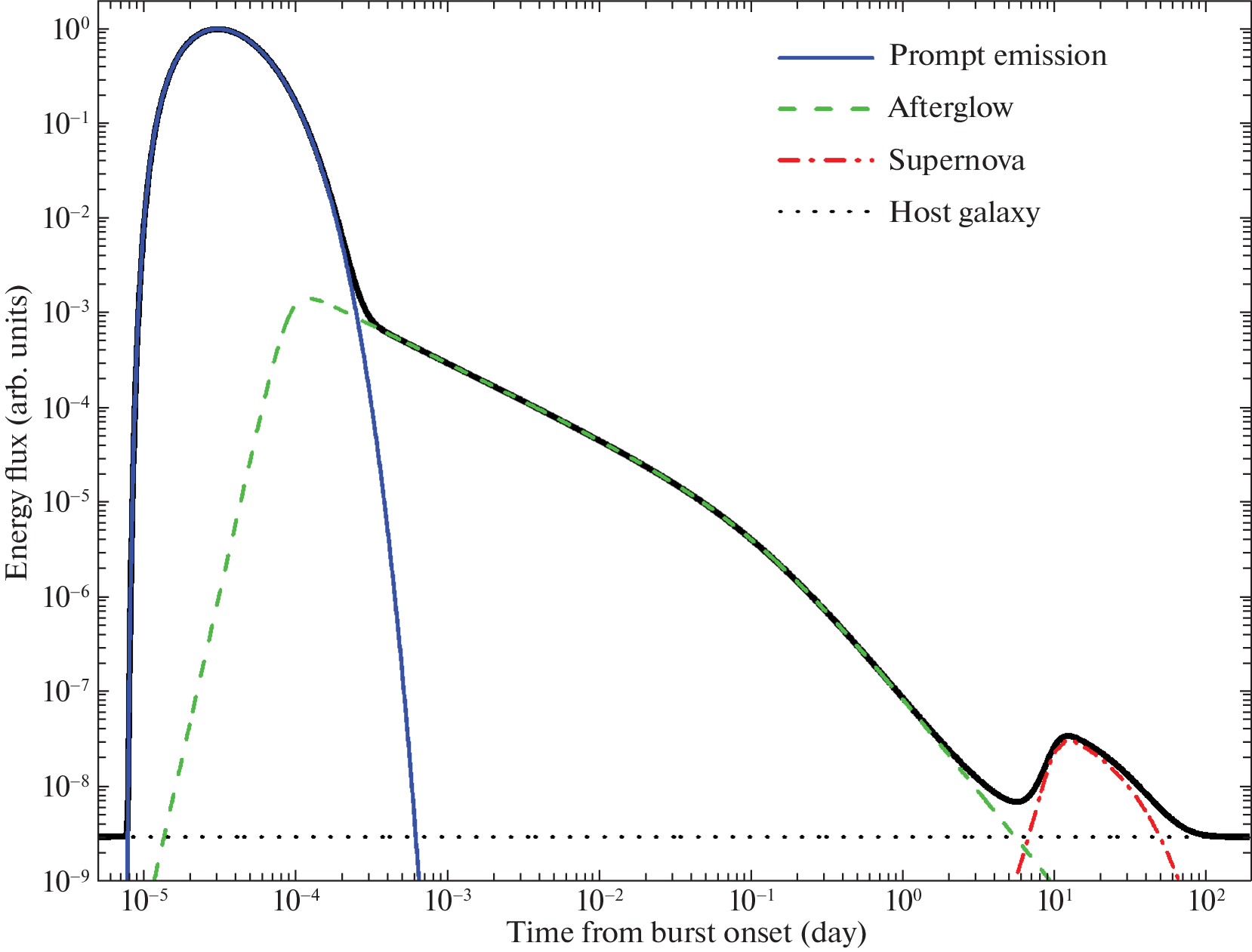}
    \caption{Different components of the optical light curve of an LGRB with an associated SN. Different contributions are labeled in the top-right corner. This figure (Figure 8 in the original paper) from~\cite{pozanenkoetal2021} was reproduced with permission from Springer Nature.}
    \label{fig:pozanenko}
\end{figure}

Furthermore, Zhang~et~al.~\cite{zhangetal2014} suggested a new method to measure the duration of a GRB as the time in which the $\gamma$- and X-rays show signatures of a relativistic jet via an internal dissipation process like late flaring activity ({see} 
 e.g.,~\cite{marguttietal2011}). By adopting this method, some GRBs can last hours; for this reason, they are classified as ultra-long GRBs (see Section~\ref{sec:ulgrbs}), but it is not yet clear if these are a truly different class of progenitors (e.g., \cite{virgilietal2013,boeretal2015}).
Although this particular physically motivated approach relies on the interpretation of X-ray flares as signatures of central engine activity, which needs to be disentangled from the contribution of the external shock (see Figure~\ref{fig:Nasa_Goddard}), it is a matter of fact that some LGRBs show a prompt phase that, if measured in X-rays (0.1--10 keV), lasts several minutes/hours before the afterglow emission becomes dominant (Figure \ref{fig:pozanenko}), challenging the standard core-collapse model (e.g.,~\cite{Woosley_1999}). The prototype of this class of ultra-long bursts is GRB~111209A, with its $\sim$7 h prompt duration in the 0.3--10 keV energy range~\cite{boeretal2015}, the longest ever measured so far. Interestingly enough, this exceptional GRB is the only one, so far, that has been found to be associated with a superluminous SN (SN~2011kl; see Section~\ref{sec:11kl}).

\section{Protagonists}
\label{sec:protagonists}
\subsection{Long and Ultra-Long GRBs}
\label{sec:grbs}
\subsubsection{GRB Phenomenology}
\label{sec:phenomenology}
In 1973, Ray W. Klebesadel, Ian B. Strong and Roy A. Olson reported on the serendipitous discovery of $\gamma$-ray detections of cosmic origin with the US Vela military satellite\endnote{The US Vela satellite was primarily used to monitor possible $\gamma$-ray emissions due to nuclear weapon tests (see, e.g.,~\cite{bonnellandklebesadel1996,viglianoandlongo2024} for reviews about the history of GRB discovery).}:
\begin{quote}
`Sixteen short bursts of photons in the energy range 0.2--1.5 MeV have been observed between 1969 July and 1972 July using widely separated spacecraft. Burst durations ranged from less than $0.1\,\mathrm{s}$ to $\sim$$30\,\mathrm{s}$, and time-integrated flux densities
from $\sim$$10^{-5}\,\mathrm{ergs\,cm^{-2}}$ to $2\times10^{-4}\,\mathrm{ergs\,cm^{-2}}$ in the energy range given.'~\cite{klebesadeletal1973}
\end{quote}
\begin{flushright}
     \footnotesize{© AAS. Reproduced with permission.}
\end{flushright}

They were initially interpreted as emissions from galactic neutron stars (NSs), but their characteristics and nearly isotropic distribution over the celestial sphere unveiled their extra-galactic origin~\cite{paczynski1986,meeganetal1992}. Between $\sim$$100\,\mathrm{keV}$ and $1\,\mathrm{MeV}$, GRBs may release about $10^{51}$--$10^{53}\,\mathrm{erg}$ (e.g.,~\cite{piran1999}) over a time scale of seconds and are observationally classified according to their duration, with spectra that are usually non-thermal \endnote{GRB spectra are usually well-fit by a smoothed broken law usually referred to as a band function~\cite{bandetal1993}.}. This evidence has constrained the ``fireball'' scenario~\cite{cavalloandrees1978,reesandmeszaros1992,meszarosandrees1993,meszarosetal1993,reesandmeszaros1994,piran1999}, in which a compact plasma with a high energy concentration (a fireball) is able to release part of its huge energy reservoir in a few seconds. In this scenario, an ultra-relativistic collimated energy outflow (a bipolar jet)~\cite{reesandmeszaros1992,piran1999} is converted into radiation. The initial energy can be either in the form of relativistic particles or emerge as a Poynting-dominated flux. Such a jet can be launched by a central compact object (see Figure~\ref{fig:Nasa_Goddard}) as a hyperaccreting black hole (BH)\endnote{A ``hyperaccreting BH'' accretes matter at an extreme rate of the order of $0.1$--$1\,\msun\,\mathrm{s}^{-1}$. In the following, we implicitly assume that BHs are stellar-mass BHs without further specification.}\cite{blandfordandznajek1977,woosley1993} or an NS~\cite{usov1992,usov1994,metzgeretal2011} via magneto-hydrodynamic (MHD) processes or via neutrino--antineutrino annihilation\endnote{It has also been shown that accretion discs are an efficient source of neutrinos and antineutrinos~\cite{eichleretal1989,pophamandwoosleyandfryer1999,ruffertandjanka1999,2001ApJ...557..949N,2002ApJ...577..311K,2004MNRAS.355..950J,chenandbeloborodov2006,2007ApJ...661.1025L,birkletal2007,zalameaandbeloborodov2011,2015ApJ...806...58L,2017NewAR..79....1L}, a fraction of which can annihilate via $\nu\overline{\nu}\rightarrow e^{-}e^{+}\rightarrow\gamma\gamma$; such a mechanism is able, in principle, to power an ultra-relativistic jet. However, Leng and Giannos~\cite{lengandgiannios2014} showed that $\nu\overline{\nu}$ annihilation does result in suitable conditions to reproduce the energy of ULGRBs ($\tn\gtrsim10^3\,\mathrm{s}$, as discussed later)
	.}. 

The need for relativistic $\Gamma$ factors is primarily to handle the so-called ``compactness problem''~\cite{piran1999} of GRBs. This apparent paradox arose in evaluating the high $\gamma$-rays to pair production optical depths ($\tau_{\gamma\gamma}$) of the order of $\tau_{\gamma\gamma}\sim10^{13}$~\cite{piran1999} for the expected radius of a GRB and the observed non-thermal GRB spectra. Goodman~\cite{goodman1986}, Paczy\'nski~\cite{paczynski1986}, and Krolik and Pier~\cite{krolikandpier1991} suggested that the radiation source moves towards the observer at relativistic velocities and that the photon frequencies are observed as blue-shifted, i.e., at the source there, are fewer photons able to produce pairs. Altogether, relativistic effects on the radius and on the optical depth reduce the average optical depth to about $\tau_{\gamma\gamma}<1$, allowing photons to escape before thermalizing. Hence, the matter outflows of a GRB must somehow be accelerated to the very high bulk velocities ($u$ or $\beta=u/c$ in units of the speed of light). The adiabatic expansion of the ejecta could, in principle, be a viable mechanism to convert the internal energy into the kinetic energy of the ejecta, a fraction of which can, in turn, be dissipated in radiation via shocks (see also Figure~\ref{fig:Nasa_Goddard}). 

The classical baryonic version of the fireball model has the advantage of simply explaining the main features observed in the GRB emission, like the prompt emission and the GRB afterglow. In particular, the prompt emission is explained by the internal-shock model, in which a central compact object launches fast matter shells following a distribution of $\Gamma$ factors, width and separations. Being supersonic, these outflow components crash into each other; their shock interaction may then generate a magnetic field due to plasma instabilities \cite
{peer2019, Huntingtonetal2017, FleishmanandUrtiev2010} and accelerate particles via synchrotron/synchrotron self-Compton~\cite{Chiabergeandghisellini1999, acciarietal2019, miceliandnava2022}. The internal shock model, though widely considered, has been questioned in some cases~\cite{zhangandyan2011}, such as that of GRB 080916C~\cite{abdoetal2009}, due to the inefficient conversion of kinetic energy to radiation caused by excessive baryonic matter in the jet, leading to the issue known as ``baryonic contamination''~\cite{piran1999}\endnote{Further details on the interpretation of GRB 080916C as an electromagnetically powered GRB and other inconsistencies in the internal shock model can be found in~\cite{zhangandyan2011}.}. When baryonic matter (such as protons and neutrons) mixes into the relativistic flow, it can significantly increase the inertia of the flow, requiring more energy to accelerate it to relativistic speeds. As a consequence, it becomes more difficult for the outflow to achieve $\Gamma>100$, which is necessary to produce the observed characteristics of GRBs and avoid the photon--photon pair production process. This latter phenomenon would otherwise reabsorb the emitted $\gamma$ rays because the energy budget is spent accelerating the baryonic mass rather than the plasma or photons. This ``baryonic contamination'' essentially hinders the ability of the outflow 
to sustain the relativistic motion required to avoid thermalization of $\gamma$ rays and to generate the observed non-thermal spectra~\cite{piran1999}. Therefore, a delicate balance is required in GRB physics to achieve the observed high-energy emissions while addressing the constraints imposed by baryonic contamination. Specifically, Piran~\cite{piran1999} showed the importance of minimizing baryon mixing for GRB outflows to maintain their relativistic nature and fit observational data.

Different models have been proposed to reduce baryon mixing. One is the photon-dominated fireball model, where the energy is predominantly carried by photons and electron--positron pairs, minimizing the impact of baryons. In this scenario, as the fireball expands and cools, it transitions into a kinetic-energy-dominated regime, reducing the impact of any residual baryonic contamination~\cite{piran1999, meszaros2001, FleishmanandUrtiev2010}. The photon-dominated fireball model provides a solid solution to the ``baryonic contamination'' problem; however, it has to be considered carefully. In fact, it has some limitations, such as in explaining the non-thermal spectral components observed in many GRBs~\cite{Peer2015} and the rapid variability observed in GRB light curves (with time scales as short as milliseconds)~\cite{BhatandGuiriec2011} and accounting for the high-energy (GeV to TeV) emissions observed in some GRBs~\cite{Gupta2007}. Because of the challenges associated with the photon-dominated model, a more promising model involves magnetically dominated outflows {and} incorporates both thermal and non-thermal processes. In this scenario, magnetic fields transport most of the energy, reducing the role of baryonic matter, thereby suppressing baryonic contamination. {This is achieved because magnetic fields are likely to be a more efficient energy source capable, in principle, of accelerating matter to very high Lorentz factors ($\Gamma=E/Mc^2\gg1$) without further enhancing the baryon loading. Another possibility is to consider the release of electromagnetic energy to launch a Poynting flux-dominated outflow
}. An important parameter describing the dynamical contribution of the magnetic energy is the magnetization parameter ($\sigma_0$)~\cite{drenkhahn2002,drenkhahnandspruit2002}, which is defined as the ratio between the Poynting flux ($L_{\rm pf}$) and the kinetic energy flux ($L_{\rm kin}$) at a {radius $r$}:
\begin{equation}
\label{eq:sigma}
\sigma_0\equiv\frac{L_{\rm pf}}{L_{\rm kin}}=\frac{\beta r^2 B^2}{4\pi\Gamma\dot{M}c}\,,
\end{equation}
where $\dot{M}$ is the mass loss rate. In the internal shock model, the magnetic field generated in the plasma\endnote{In this case, $\sigma_0$ is different from Equation (\ref{eq:sigma})~\cite{zhangandyan2011}.} is not dynamically very important, and $\sigma_0\ll1$. On the contrary, in a Poynting-dominated flux, $\sigma_0\gg1$, and the prompt emission is caused by the dissipation of magnetic energy via turbulent magnetic reconnections~\cite{drenkhahnandspruit2002,lyutikovandblandford2003}\endnote{This scenario was also recently applied to GRB~230307A~\cite{duetal2024}.}. In addition to these two models, collimated outflows have also been proposed as a solution to baryonic contamination. These outflows should focus the relativistic jet into a narrow angle, effectively ``cleaning'' the jet as it interacts with surrounding material~\cite{piran1999, Zhangandmeszaros2002, FleishmanandUrtiev2010}.

The schematic presented in Figure~\ref{fig:Nasa_Goddard} shows the evolution of a GRB through internal and external shocks. When the relativistic jet is launched and propagates through a massive progenitor star, trying to break through its outer layers, it interacts with the stellar envelope, inducing shocks that dissipate part of its energy by heating and compressing the surrounding gas, forming a so-called {``}cocoon{''} ~\cite{ramirezruizetal2002,lazzatietal2005}. This cocoon consists of high-pressure, hot plasma formed from the shocked stellar material and jet material. It plays a crucial role in shaping the structure of the jet by exerting pressure on it, which also alters its dynamics and angular energy distribution~\cite{GottliebandNakarandPiran2018,eisenbergandgottliebandnakar2022}. As a result, the jet develops a structured outflow, with a narrow core of a high Lorentz factor and a broader region of slower material. This structure helps explain the different properties of GRBs observed from different viewing angles. The cocoon itself stores a significant amount of energy and can escape the progenitor star in either an isotropic or anisotropic manner, depending on the structure of the star and the characteristics of the jet. This energy release may contribute to various observational phenomena, such as low-luminosity GRBs (LLGRBs) or X-ray-rich transients~\cite{eisenbergandgottliebandnakar2022}. 
Therefore, the presence of a cocoon is key to explaining some of the diversity observed in GRBs and associated SNe~\cite{LazzatiandMorsonyandBegelman2010}. Their production has also been extensively investigated~\cite{LazzatiandMorsonyandBegelman2010, GottliebandNakarandPiran2018}. There are several ways in which the cocoon can alter the observed properties of the event~\cite{GottliebandNakarandPiran2018, LazzatiandMorsonyandBegelman2010, Brombergetal2011, nakarandpiran2017}. 
If the jet successfully breaks out of the star, the primary GRB prompt emission occurs, often followed by a multi-wavelength afterglow. The cocoon may also break out, contributing to a softer isotropic emission (lower-energy $\gamma$ rays) during the afterglow, which is often observed in conjunction with the high-energy $\gamma$-ray signal. Furthermore, the combination of the emission of the collimated jet and the typically broader and softer emission of the cocoon might explain the broad-band spectra seen in some GRBs~\cite{GottliebandNakarandPiran2018, nakarandpiran2017}. This allows for the prediction of specific afterglow features that can be used to distinguish between jet-dominated and cocoon-dominated events.
Conversely, if the jet fails to break out due to insufficient energy or excessive stellar material, the cocoon may still escape~\cite{nakarandpiran2017, LazzatiandMorsonyandBegelman2010}. In such cases, an orphan GRB\endnote{An ``orphan'' GRB is a GRB event in which the typical high-energy $\gamma$-ray emission is absent or undetected but other afterglow signatures (such as in optical, X-ray or radio wavelengths) are still observed.} may result, or the event may manifest as an SN Ic BL without the characteristic narrow beam of a GRB. These are sometimes referred to as ``failed GRBs'' or ``cocoon-driven explosions''. This process might also explain LLGRBs, which lack a clear central jet~\cite{Brombergetal2011, GottliebandNakarandPiran2018}.
\subsubsection{Ultra Long GRBs}
\label{sec:ulgrbs}
\begin{figure}[H]
    
    \includegraphics[width=0.8\linewidth]{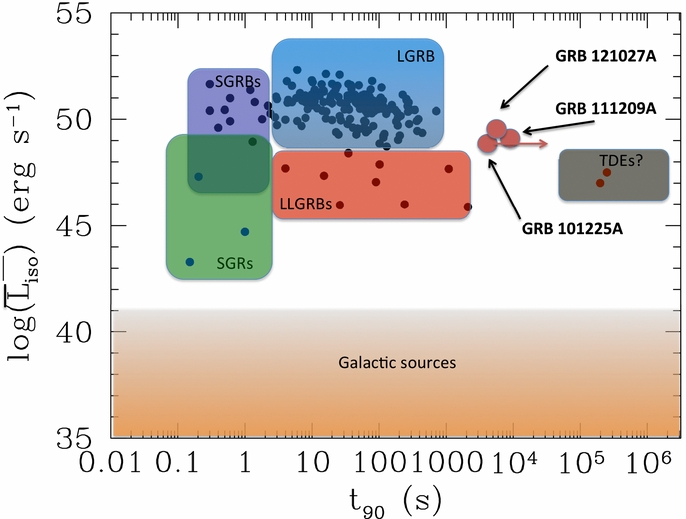}
    \caption{Distribution of different classes of GRBs according to their typical luminosity and duration. The special cases of GRB 111209A, 121027A and 101225A (see Section~\ref{sec:pecgrbsne}) are highlighted among a sample of ULGRBs, SGRBs, LGRBs, low-luminosity GRBs (LLGRBs in the figure)~\cite{liangetal2007}, soft gamma repeaters (SGRs in the figure)~\cite{woodsandthompson2006,mereghetti2008} and tidal disruption events (TDEs). This figure is from~\cite{levanetal2014} (Figure~2 in the original paper). © AAS. Reproduced with permission.}
    \label{fig:ulgrbs}
\end{figure}

It is often the case in astronomy that even the best established taxonomy is likely to be widened by future observations. GRBs make no exception: together with some intermediate-duration GRBs lasting 2--5 s~\cite{mukherjeeetal1998,tunnicliffeandlevan2012} or the so-called ``extended-emission'' of some short GRBs (e.g.,~\cite{norrisandbonnell2006,rastinejadetal2022,yangetal2022}), very long-lasting GRBs have been discovered to populate the end of the GRB duration distribution and are therefore referred to as ultra long gamma-ray bursts (ULGRBs)~\cite{connaughton2002,tikhomirovaandstern2005,gruberetal2011,thoeneetal2011b,gendreetal2013,strattaetal2013,virgilietal2013,evansetal2014,levanetal2014,piroetal2014} (see Figure~\ref{fig:ulgrbs}). Whether or not the latter corresponds to a distinct category of objects is still debated~\cite{virgilietal2013,levanetal2014,boeretal2015}, as, above all, it is difficult to establish a precise threshold in the $\tn$ space to discriminate between LGRBs and ULGRBs (see also Section~\ref{sec:duration}). In fact, this approach has some biases: (i)
 The measurement of the GRB duration is complicated by occultation by the Earth, the orbital parameters, the sensitivity and the energy band of the instrument and the discontinuous emission of some objects~\cite{roretal2024}. (ii) The GRB signal should ideally be disentangled by external contributions that are unrelated to the GRB itself. Furthermore, some unprecedented observations of GRBs have led to ambiguity regarding whether their duration is sufficient to comprehensively characterize an event as a collapsar or a compact merger. Different attempts have been suggested by different authors (e.g.,~\cite{campanaetal2006,starlingetal2011,levanetal2014,zhangetal2014,greineretal2015}), but their nature and the need to isolate a new GRB population remain elusive.

Recently, Ror~et~al.~\cite{roretal2024} analyzed the exceptional case of ULGRB GRB 221009A, which is not only the brightest ULGRB ever observed but also the only one associated with very high-energy (GeV-TeV) detections. The authors compared the prompt and afterglow properties of GRB 221009A with those of a broad sample of \textit{Swift}-detected GRBs with known redshift. 
Throughout their analysis, they adopted the duration as the main discriminant between the different GRB types. To do this, the authors built a log-normal distribution of the GRB sample against their $\tn$ and isolated the LGRB component, which was consequently fitted with a Gaussian. LGRBs with $\tn>\langle\tn\rangle=43\,\mathrm{s}$ were divided into three 1-sigma-wide bins, respectively referred to as bronze, silver and gold\endnote{The authors also mentioned a diamond subsample made by a number of very well known ULGRBs, which, due to their $\tn$, entirely fall within the gold subsample.} subsamples (see Figure~2 in~\cite{roretal2024}). With $\tn\gtrsim1000$ s, GRB 221009A was included in the gold subsample. In addition, the \textit{Swift}
GRBs were subdivided according to the morphological properties of their emissions via a machine learning tool, but the authors disfavored this approach to distinguish between different GRB progenitors. They also identified trends in the spectral parameters within the same sample and found that the gold subsample exhibited a lower average hardness ratio compared to the silver and bronze subsampled and that the fluence of the gold and the silver subsamples versus the $\tn$ does not obey to the increasing trend of the whole GRB sample (see Figure 6 therein, upper panel). Finally, Ror~et~al. investigated LGRB scenarios, comparing the total isotropic kinetic and $\gamma$ energies with the maximum rotational energy achieved by a ``magnetar'', i.e., a highly magnetized and rapidly rotating NS with a magnetic field of the order of $\sim$$10^{14}$--$10^{15}$~G (far exceeding that of a typical NS of $\sim$$10^8$--$10^{12}$~G)~\cite{thompsonandduncan1993,Spruit2008} and millisecond spin periods. Given the magnetar maximum spin energy  ($\sim$$2\times10^{52}\,\mathrm{erg}$), Ror~et~al. found that three out of twenty-two ULGRBs belonging to the gold subsample strongly favored a magnetar progenitor. Altogether, the properties of the bronze subsample did not significantly differ from those of LGRBs with $\tn<\langle\tn\rangle$, while the gold subsample could potentially represent a distinct GRB population. This result corroborates past findings on the GRB prompt emission distribution in the soft X-ray energy range~\cite{boeretal2015}.

Besides a possible magnetar origin, ULGRBs have been proposed to be linked to tidal disruption events or to extended massive stars, such as blue super giants (BSGs). This is the case, for instance, for ULGRB GRB 111209A, for which a BSG progenitor has been suggested~\cite{gendreetal2013}. BSGs could potentially explain the ultra-long duration of ULGRBs because of the expected long free-fall time scales of their envelopes, which allow accretion to power a long-lasting central engine. Previous results based on numerical simulations~\cite{pernaetal2018} showed that a jet can emerge from a BSG, and the resulting light curves resemble those observed in ULGRBs, with durations ranging from $O(10^3)$ to $O(10^4)$ seconds, in accordance with the observations. However, as noted by Kann~et~al.~\cite{Kann2019}, while BSGs are able to produce ULGRBs, no or faint SNe are expected in this scenario, in contrast to the observation of an SLSN associated with GRB~111009A (see also Section~\ref{sec:discussion}). 
\subsection{Classical and Superluminous Supernovae}
SNe are among the most studied objects in the history of scientific astronomy~\cite{aldallalandazzam2021}.
The first known observations of an SN date back to 185 CE and were reported by Chinese astronomers~\cite{zhaoetal2006} who observed an emerging ``guest star'' near $\alpha\,$Cen:
\begin{quote}
    `It seemed to be as large as half a yan, with scintillating, variegated colors, and it then grew smaller, until in the sixth month of the \textit{hou}-year (\textit{hou-nian}, 24 July to 23 August AD 187), it disappeared.'~\cite{zhaoetal2006}
\end{quote}
\begin{flushright}
     \footnotesize{© AAS. Reproduced with permission.}
\end{flushright}

At the time, the only means of observation was the naked eye, which was limited to the optical light emitted by galactic events. After about 2000 years, astronomers developed techniques to follow the evolution of many SNe, even far outside the Milky Way and in a wider electromagnetic range. This increased the number of SN discoveries and the characterization of different SN subtypes (see Section~\ref{sec:snzoo}) and deepened our knowledge about their explosion mechanisms and central engines~\mbox{\cite{macfadyenandwoosley1999,janka2012,mulleretal2012,burrows2013,Betheandwilson1985,Mulleretal2013,mezzacappaandbruenn1999,Metzgeretal2008, dessartetal2008,bucciantinietal2012,shankaretal2021,lazzatietal2005,barnesetal2018,eisenbergandgottliebandnakar2022,pophamandwoosleyandfryer1999,Janiukandmoderskiandproga2008, Menegazzietal2024a, Menegazzietal2024b,obergaulingerandaloy2017,obergaulingerandaloy2020, radiceetal2019}}. Still, this is not all we need to completely unveil their nature, in particular for the SN subtypes that were recently discovered by the new generation, including wide-field and large etendue transient surveys~\cite{ivezicetal2019,foersteretal2021}
.
\subsubsection{The Supernova Zoo}
\label{sec:snzoo}
\begin{figure}[H]
    
    \includegraphics[width=0.8\linewidth]{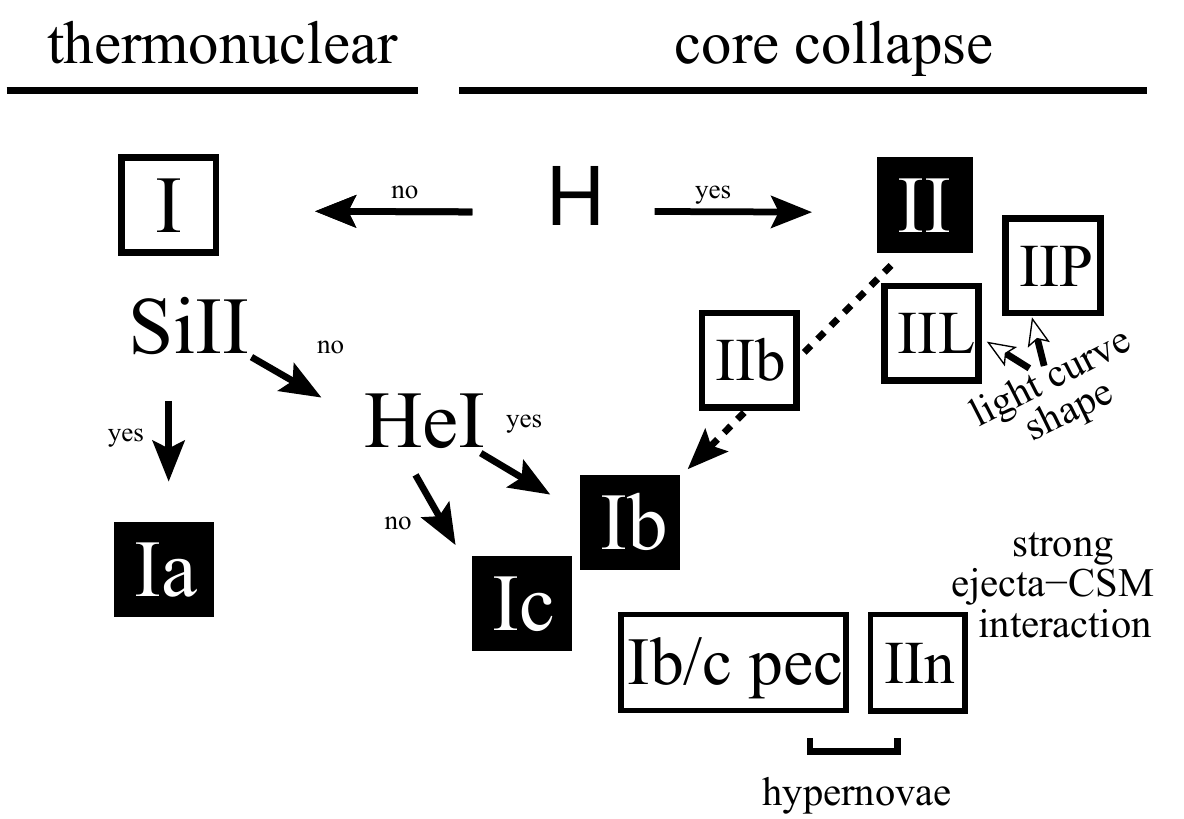}
    \caption{SN classification tree distinguishing between thermonuclear and core-collapse SN explosions. This figure from~\cite{turatto2003} (Figure 1 in the original paper) was reproduced with permission from Springer Nature. }
    \label{fig:taxonomy}
\end{figure}

As SNe reflect the different characteristics of their progenitor stars, their light curves and spectra can vary considerably in terms of chemical composition, brightness and duration, and they are usually classified by looking at an optical spectrum taken at about the maximum luminosity, in which the absence or presence of Balmer lines in their light curve discriminates between type I and type II SNe, respectively
. In detail, type I SNe are referred to as SNe Ia if they show spectral silicon abundance, SNe Ib if they show spectral helium abundance or SNe Ic if they show neither of the two. Type II SNe are distinguished between type II P and type II L depending on whether the initial part of the light curve has a ``plateau'' (P) or a ``linear declining'' (L) phase after maximum luminosity (see also Figure~\ref{fig:taxonomy}). 
SNe Ia are interpreted as a thermonuclear explosion of a white dwarf in a close binary system~\cite{nomoto1986, bravoetal2024, blondin2024}, while all other types 
are usually interpreted as a core-collapse explosion~\cite{woosleyandheger2006, jankaetal2016, burrowsandvartanyan2021, bolligetal2021, mezzacappaetal2020, Vartanyanandcolemanandburrows2021, kurodaetal2022, Bruennetal2023, Menegazzietal2024a, Menegazzietal2024b}. However, this scheme has been further expanded to encompass more peculiar SNe whose classifications do not comfortably suit the above-sketched classification algorithm; this is the case of the interacting SNe (type IIn and type Ibn SNe, e.g.,~\cite{schlegel1990,filippenko1997,smith2017}), the various peculiar types of SNe Ia (e.g.,~\cite{taubenberger2017}) and SNe Ic, whose lines are remarkably broader compared to those of standard SNe, i.e., SNe Ic BL, to name a few. SNe Ic BL are often spatially and temporally associated with LGRBs. It is important to stress that ``hypernova'' and ``SN Ic BL'' are not strictly synonyms; the latter is an observational classification grouping SNe Ic, whose remarkable line broadening implies bulk ejecta velocities of $\sim$15,000--30,000$\,\mathrm{km\,s^{-1}}$ (e.g.,~\cite{modjazetal2016,taddiaetal2019}), while ``hypernova'' identifies very energetic SNe, most of which are associated with LGRBs and, in this case, termed GRB~SNe\endnote{The use of this terminology is not standard in astronomy and might be misleading. For instance, the terms ``hypernovae'' and GRB~SNe have also been used differently, e.g., in the case of SN Ic SN~1997ef~\cite{iwamotoetal2000}, which was not associated with any LGRB observation but referred to as a hypernova. Moreover, the definition of ``SN Ic BL'' is actually somewhat arbitrary without a precise broadness threshold (see~\cite{prenticeandmazzali2017}).}. 

There could be several reasons to explain why not every SN Ic BL/hypernova is observed in association with an LGRB. One possibility considers that the relativistic jet associated with the GRB is highly collimated and does not point toward Earth, making the GRB undetectable and leaving the SN signature only. In addition, observational limitations, such as intrinsic faintness or extreme distance, may cause the GRB signal to fall below the detection threshold of current $\gamma$-ray telescopes. Another explanation could be that the relativistic jets in these events have a low Lorentz factor or insufficient energy to generate detectable $\gamma$ rays, even though the explosion mechanism is analogous to GRB-associated events~\cite{shankaretal2021}. Far less easy to explain are LGRBs for which no hypernova companion can be detected. An interesting hypothesis considers that those GRBs, even if lasting more than 2 s, do not arise from a collapsar event but from a compact binary merger and, hence, are accompanied by much fainter electromagnetic counterparts~\cite{yangetal2022}. This hypothesis is also fostered by their typical host galaxies and redshift distributions~\cite{vanputtenetal2014,taggartandperley2021}.
\subsubsection{Superluminous Supernovae}
\label{sec:slsne}
SLSNe~\cite{richardsonetal2002,galyam2012,howell2017,galyam2019} are a class of exotic transients discovered in the last two decades. As their name hints, SLSNe are exceptionally luminous SNe, as their magnitudes can be as bright as $\approx$$-20$ mag or even more at maximum luminosity. Initially, an absolute magnitude at maximum in optical bands of $-21$ mag was set to separate SLSNe from their standard siblings~\cite{galyam2012}, but, as for the latter, it is crucial to observe an optical spectrum at pre-maximum/maximum phases to properly classify SLSNe. This first allows for a sub-classification between hydrogen-devoid and hydrogen-rich events, termed SLSNe I and SLSNe II, respectively. Secondly, if we restrict the discussion to the case of SLSNe I, prominent P-Cygni absorptions between $\sim$3500 and 4500 \AA{} usually interpreted as transitions from a highly excited level of the \ion{O}{ii} (e.g.,~\cite{mazzalietal2016,quimbyetal2018}) \endnote{See~\cite{koenyvestoth2022} for a different interpretation.} are usually seen in SLSNe I spectra at pre-maximum/maximum and are an almost unique feature of the great majority of SLSNe I\endnote{See the cases of type Ib SN~2008D~\cite{soderbergetal2008} showing the \ion{O}{ii} lines and SLSN I SN~2020wnt~\cite{gutierrezetal2022}, which does not.}. In the luminosity space, there is no evidence for a bimodal distribution between stripped-envelope CCSNe and SLSNe I~\cite{prenticeetal2021,gomezetal2022}. Within SLSNe II, there is also another subclass termed SLSNe IIn. Such SLSNe do not really pose a challenge, since, similar to SNe IIn, their spectra usually show multi-component/narrow Balmer features, a signature that unambiguously indicates the interaction of the SN ejecta with low-velocity material lost by the progenitor star prior to the SN explosion, as discussed later in this section. 

With the discovery of SN~2011kl~\cite{greineretal2015}, the association between an LGRB and an SLSN I was unprecedented but, due to two strong connections between SLSNe I and SNe Ic, foreseeable. In fact, while SLSNe spectra at the maximum are very hot (usually with a black-body temperature $10^4\,\mathrm{K}\leq T_{\rm bb}\leq10^5\,\mathrm{K}$) and almost featureless, after $\sim$$15$--$30$ days from maximum luminosity, the spectrum of an SLSN I enters a new phase in which it faithfully reproduces the main features of an SN Ic (eventually in their BL fashion) at its maximum luminosity  (see, e.g.,~\cite{pastorelloetal2010}). In fact, from this point onward, SLSNe I spectra also show some features from singly ionized ions like \ion{Mg}{ii}, \ion{Fe}{ii} and \ion{Si}{ii} (see, e.g.,~\cite{howell2017}). Secondly, galaxies hosting SLSNe I and hypernovae (hence, LGRBs) share similar properties and candidate scenarios to be modeled (see Section~\ref{sec:environments}). For these reasons, we mostly deal with SLSNe~I, as they are of major interest for the present review.

Light curves of SLSNe I can be very complex and diverse. They evolve over a very broad range of time scales: they usually rise to the maximum luminosity in \mbox{$\tau\approx20$--$50$}~rest-frame days~\cite{inserra2019} and decline in $f\times\tau$ rest-frame days, where $f\approx1$--$2$~\cite{nicholletal2015}. It was initially suggested that SLSNe I were intrinsically divided between fast- and slow-evolving objects, but this was subsequently disfavored by studies considering wider samples~\cite{nicholletal2015,deciaetal2018,lunnanetal2018,angusetal2019} and objects with intermediate evolutionary rates~\cite{kangasetal2017,fioreetal2021}. In addition, light curves of many SLSNe I are characterized by bumps and/or modulations both before and after the maximum luminosity (e.g.,~\cite{nicholletal2015,yanetal2015,yanetal2017,pursianenetal2022,westetal2023,zhuetal2023,zhuetal2024}), and some of them have shown a ``knee'' after which the light curve settles on an exponential tail~\cite{inserraetal2013}, somehow resembling the radioactive tails foreseen by the classical $^{56}$Ni decay-driven SN scenario, according to which
 explosive silicon burning produces $^{56}$Ni, which, via electron capture and positron decay~\cite{nadyozhin1994}, in turn, decays in $^{56}$Co and $^{56}$Fe as follows:
\begin{equation}
    ^{56}\mathrm{Ni}\rightarrow ^{56}\mathrm{Co}\rightarrow^{56}\mathrm{Fe}+e^- + \gamma + \nu_e\,.
\end{equation}

The abundance $N$ of $^{56}$Ni and $^{56}$Co decays is regulated by exponential decay laws like the following:
\begin{equation}
\label{eq:radioactive1}
\frac{\ud N}{\ud t}=-\lambda N\,,
\end{equation}
where $\lambda=\ln 2/t_{1/2}$ is the decay rate per unit time and $t_{1/2}$ is the half life of the isotope. For $^{56}$Ni and $^{56}$Co, $t_{1/2}(^{56}\mathrm{Ni})=6.10$ days and $t_{1/2}(^{56}\mathrm{Co})=77.12$ days, respectively~\cite{nadyozhin1994}. Assuming that the radiation flux should be proportional to $\ud N/\ud t$, converting flux in bolometric magnitudes ($M_{\rm bol}$) via the Pogson law and assuming full trapping of electrons and $\gamma$ rays, Equation (\ref{eq:radioactive1}) can be written as follows (see also~\cite{padmanabhan2001}):
\begin{align}
    &\frac{\ud M_{\rm bol}(^{56}\mathrm{Ni})}{\ud t}\simeq0.12\,\mathrm{mag\,d^{-1}}\,,\\
    &\frac{\ud M_{\rm bol}(^{56}\mathrm{Co})}{\ud t}\simeq0.0098\,\mathrm{mag\,d^{-1}}\,.
\end{align}

These slopes can often be directly observed in the bolometric light curve of SNe. Simulations show that core-collapse explosions produce $\sim$$0.04$--$0.2$ $\msun$ of $^{56}$Ni (e.g.,~\cite{anderson2019,uglianoetal2012,Menegazzietal2024a, Menegazzietal2024b}) depending on the SN type. If one applies this scenario to the case of an SLSN, the predicted $^{56}$Ni masses are of the order of $1-5\,\msun$. Given the typical energies ($\lesssim$$5\times10^{52}\,\mathrm{erg}$ of a $\nu$-driven core collapse, synthesizing such a huge amount of $^{56}$Ni can be challenging for standard core-collapse SN models. Much higher energies can be reached by a pair-instability SN scenario in which the explosion of a very massive progenitor (\mbox{$\sim$$140$--$260\,\mathrm{M_\odot}$}~\cite{hegerandwoosley2002}) can meet the thermodynamic conditions necessary to trigger $e^+,e^-$-pair creation. This process leads to a sudden softening of the equation of state, letting the star collapse and explode in a violent thermonuclear runaway. While the pair-instability scenario is, in principle, a viable mechanism to produce more $^{56}$Ni, the high ejecta masses and opacities expected from a pair-instability SN make its light curve very slow-rising~\cite{nicholletal2013} (see~\mbox{\cite{kozyrevaandblinnikov2015,mazzalietal2019,moriyaetal2019}}) and its spectra suppressed on the UV/blue side~\cite{dessartetal2013,jerkstrandandsmarttandheger2016}, respectively.

\textls[-15]{As an alternative to $^{56}$Ni decay, a different mechanism can be invoked to power SLSNe. Several hypotheses have been put forward to explain them, like the ejecta--circumstellar material interaction}~\cite{chevalierandfransson2003,chevalierandirwin2011,chatzopoulosetal2012,ginzburgandbalberg2012,nicholletal2014,chenetal2015}, fallback accretion~\cite{dexterandkasen2013,kasenandmetzgerandbildsten2016,moriyaandnichollandguillochon2018} and the thermalization of a photon-pair plasma nebula inflated by the spin-down radiation of a millisecond magnetar (see Section~\ref{sec:magnetarsn}). Despite many studies searching for, e.g., polarimetry signatures~\mbox{\cite{cikotaetal2018,lee2020,poidevinetal2022,pursiainenetal2023}} or hunting for the high-energy signals of a compact remnant~\cite{fioreetal2022,lietal2024}, the lack of clear independent evidence for one of the proposed scenarios makes consensus more difficult to be reached with respect to the nature of SLSNe, and it is most likely that more than one mechanism is at play for some events. This could be particularly true for SLSNe I displaying light-curve modulations before and/or after the maximum-luminosity epoch. These can be suitably explained as an additional ejecta--circumstellar material (CSM) interaction,  
which is possibly unrelated to the mechanism responsible of the peak luminosity. Pre-explosion mass loss cold be due, e.g., to wind mass loss or to the so-called pulsational pair instability (PPI) phenomenon (e.g.,~\cite{woosleyandblinnikovandheger2007,woosley2017,renzoetal2020}). Unlike pair-instability SNe, PPI is not enough to unbind the progenitor star but can eject a sizable amount of material prior to the SN explosion
. This phenomenon places some shells of matter in front of the star, which the ejecta shocks after the SN explosion. In this scenario, such an interaction with each shell corresponds to a single bump in the light curve. While this scenario is fully reasonable from an astrophysical point of view, it is challenging to explain the lack of narrow/multicomponent\endnote{Narrow/multi-component features are a clear spectroscopic signature of CSM interaction, and they are usually seen in the spectra of SNe IIn~\cite{smith2017}.} H/He features in the spectra of SLSNe I. Andrews and Smith~\cite{andrewsandsmith2018} suggested that these features can be masked by a peculiar geometry of the ejecta. Another possibility considers that at least some SLSNe~I, as well as SNe Ic BL, might be associated with magnetorotationally driven core-collapse explosions~\cite{leblancandwilson1970,bisnovatyiandpopovandsamokhin1976,meieretal1976,muellerandhillebrandt1979}, some of which are able to produce both a sizable amount of $^{56}$Ni and a magnetar remnant. This does not preclude the possible association of magnetorotationally driven explosions with LGRBs~\cite{reichertetal2023}.

It is worth mentioning that the study of SLSNe has the potential to significantly impact contemporary astrophysics. {In addition to challenging stellar evolution and transient astronomy, their bright luminosities make them of interest for cosmology as well.} In fact, a number of studies~\cite{inserraandsmartt2014,weietal2015,inserraetal2021,khetanandcookeandbranchesi2023} suggest their possible use as high-$z$ distance indicators, potentially extending the Hubble diagram beyond $z>1.5$.
\section{Magnetars}
\label{sec:magnetars}
A growing number of astrophysical objects have been putatively interpreted with the contribution of NSs. Due to their flexibility in modeling even very diverse and often exotic phenomena, they deserve the reputation of a ``patch'' to unravel many otherwise unsolvable riddles of modern astrophysics, like soft gamma repeaters~\cite{thompsonandduncan1996}; ultraluminous X-ray pulsars~\cite{bachettietal2014}; fast radio bursts~\cite{bochenek2020}; and, not least, SLSNe and GRBs. The link between massive, non-degenerate stars and NSs was very well established by classical stellar evolution theories, and invoking their contribution is fully reasonable from an astrophysical perspective. Magnetars are highly magnetized NSs, which means that their magnetic-field strength ($B$) exceeds the quantum critical field ($B_{\rm QED}={m_e^2c^3}/{e\hbar}\simeq 4.4\times10^{13}\,\mathrm{G}$, where $m_e$ is the electron mass, $c$ is the speed of light in vacuum and $\hbar$ is the reduced Planck constant). This value corresponds to the magnetic field for which the cyclotron energy equals the rest of the energy of the electron. Even if created during the core collapse of a massive star, such magnetic-field values likely require the invocation of a magnetic-field amplification mechanism instead of more conventionally assuming magnetic flux conservation during the core collapse for typical values of the progenitor magnetic field\endnote{Typical values of the initial magnetic field at the beginning of the core collapse are of the order of $10^2\,\mathrm{G}$~\cite{aurieretal2003,huetal2009} inherited, e.g., as a ``fossil field'' by the star-formation process~\cite{braithwaiteandspruit2004}.}. Possible $B$-field amplification mechanisms consider the formation of a proto-NS (PNS) with periods between $\sim$$0.6$~ms and 3~ms~\cite{thompsonandduncan1993}. In this scenario, the strong magnetic field experienced by magnetars is thought to result from an amplification process caused by convection and differential rotation in such rapidly rotating stars (e.g.,~\cite{duncanandthompson1992,thompsonandduncan1993, FryerandWarren2004, Burasetal2006}). 

The high magnetic field of the magnetar leads to extremely intense electromagnetic radiation, and when coupled with the NS rotational energy, this can generate powerful jets. These jets are thought to be one of the possible mechanisms responsible for the relativistic outflows observed in GRBs (e.g.,~\cite{metzgeretal2011}). In a simplified fashion, a magnetar with an inertia moment ($I$) and an angular velocity ($\Omega$) looses its rotational energy as
\begin{equation}
\label{eq:rotenergy}
E_{\rm rot}=\frac{1}{2}I\Omega^2\simeq2\times10^{52}\,M_{1.4}R_6^2P_{-3}^2\
\end{equation}
via magnetic-dipole radiation ($\dot{E}_{\rm md}$) and gravitational-quadrupole radiation ($\dot{E}_{\rm gq}$) (see e.g.,~\cite{ostrikerandgunn1969}):
\begin{linenomath}
    \begin{align}
        &\dot{E}_{\rm md}=-\frac{2}{3}\mu_\perp^2\,,\\
        &\dot{E}_{\rm gq}=-\frac{1}{45}GD_\perp^2\frac{\Omega^6}{c^5}\,.
    \end{align}
\end{linenomath}

Hence, one can impose the following:
\begin{linenomath}
    \begin{equation}
        \label{eq:spindown0}
        \dot{E}_{\rm rot}=\dot{E}_{\rm md}+\dot{E}_{\rm gq}\,,
    \end{equation}
\end{linenomath}
where $\mu_\perp=BR^2\sin\Theta$ is the perpendicular component of the magnetic dipole moment, $G$ is the gravitational constant and $D_\perp$ is the mass quadrupole. Neglecting the contribution of $\dot{E}_{\rm gq}$ for simplicity, Equation~(\ref{eq:spindown0}) reads as follows:
\begin{linenomath}
    \begin{equation}
    \label{eq:spindown1}
        -\dot{E}_{\rm rot}=-\frac{\ud}{\ud t}\left(\frac{1}{2}I\Omega^2\right)=\frac{2}{3c^3}\mu_\perp^2=\frac{2}{3c^3}B^2R^6\Omega^4\sin^2{\Theta}\,,
    \end{equation}
\end{linenomath}
where $\Theta$ is the phase angle between the directions of the NS angular momentum. Assuming $I$ is independent of time, Equation~(\ref{eq:spindown1}) reads as follows:

\begin{linenomath}
    \begin{equation}
        \label{eq:spindown2}
        \dot{\Omega}=-\left(\frac{2B^2R^6}{3Ic^3}\sin^2{\Theta}\right)\Omega^3\,.
    \end{equation}
\end{linenomath}

Integrating Equation~(\ref{eq:spindown2}) between $t$ and 0 (corresponding to angular velocities $\Omega(t)$ and $\Omega_0$, respectively) results in the following:
\begin{linenomath}
    \begin{equation}
        \label{eq:spindown3}
        \Omega(t)=\frac{\Omega_0}{(1+t/t_{\rm sd})^{1/2}}\,,
    \end{equation}
\end{linenomath}
where we assume $\sin^2{\Theta}=0.5$ (e.g.,~\cite{kasenandbildsten2010}) and define the spin-down time scale ($t_{\rm sd}$)~as
\begin{linenomath}
    \begin{equation}
        \label{eq:sdtime}
        t_{\rm sd}=\frac{6Ic^3}{B^2R^6\Omega_0^2}\simeq4.1\times10^5\, B_{14}^{-2} P_{0,1}^{2}\quad\mathrm{s}\,,
    \end{equation}
\end{linenomath}
where $P_0=2\pi/\Omega_0$ is the initial period. An expression for the luminosity due to rotational energy losses can be obtained by substituting Equation~(\ref{eq:spindown3}) into Equation~(\ref{eq:spindown1}) as follows:
\begin{linenomath}
    \begin{equation}
        \label{eq:sdlum1}
        L_{\rm sd}(t)=\frac{L_0}{(1+t/t_{\rm sd})^2}\,,
    \end{equation}
\end{linenomath}
where
\begin{linenomath}
    \begin{equation}
    \label{eq:sdlum2}
        L_0=\frac{B^2R^6\Omega_0^4}{3c^3}\simeq4.9\times10^{46}\, B_{14}^{2} P_{0,1}^{-4}\,\mathrm{erg\,s^{-1}}\,.
    \end{equation}
\end{linenomath}

It is then licit to ask in which environments magnetars usually form. Unlike BH formation models, which require low-metallicity environments (with metallicities typically below $0.3$--$0.5\,Z_\odot$~\cite{fruchteretal2006, modjazetal2008}) to reduce the angular momentum losses due to stellar wind~\cite{Mandelanddemink2016, Vinkandharries2017, dicarloetal2020}, magnetar progenitors are less constrained by metallicity, allowing for their formation across a wider range of stellar environments. Low-metallicity galaxies are still the favored candidates for the formation of magnetars, since they promote the formation of rapidly rotating massive stars by reducing stellar wind, allowing massive stars to retain more angular momentum during their evolution, a necessary condition for magnetar scenarios~\cite{yoonandlanger2005, woosleyandheger2006}. However, recent studies have explored the formation of magnetars in metal-rich environments, suggesting that they can still support the creation of magnetars under specific conditions. For instance, research indicates that in metal-rich galaxies, higher metallicity leads to increased stellar mass loss, which can result in the merging of stars. This merger process may produce more massive stars with stronger magnetic fields, potentially leading to the formation of magnetars~\cite{Ablimitetal2022, sharmaetal2024}. Another possible situation for the formation of magnetars in metal-rich galaxies is when the progenitor stars are sufficiently massive and retain enough angular momentum~\cite{Songandliu2023}.
 This flexibility is a key feature that makes magnetars able to account for GRBs and SNe in both metal-rich and metal-poor galaxies (e.g.,~\cite{modjazetal2016, kasenandmetzgerandbildsten2016, Songandliu2023, sharmaetal2024}).
\subsection{Magnetar-Powered SNe}
\label{sec:magnetarsn}
The idea that an NS could contribute to the dynamics and/or the luminosity of an SN is widely used in the SN literature, but it is not novel. Based on the pioneering findings of Baade and Zwicky and Bodenheimer and Ostriker~\cite{bodenheimerandostriker1974}, Ostriker and Gunn~\cite{ostrikerandgunn1971} suggested that a central pulsar may energize the mass ejected during an SN event at the expense of its magnetic dipole luminosity, while Gaffett~\cite{gaffett1977a,gaffett1977b} investigated effects on SN light curves considering the radiative heat-diffusion problem assuming a uniform-density medium for the surrounding gaseous envelope. Observations of very luminous and energetic SNe questioned the role of $^{56}\mathrm{Ni}$ decay as their main heating source in order to avoid the requirement of unreasonably high masses of $^{56}\mathrm{Ni}$ to be interpreted. This was suggested by Folatelli~et~al.~\cite{folatellietal2006} to explain the peculiar SN~2005bf, a very energetic SN with a slow rise ($\sim$$40$ rest-frame days) towards maximum luminosity and a double-peaked light curve and spectra showing possible signatures of ejecta anisotropies\endnote{A possible signature revealing a departure from the spherical symmetry of the ejecta is the double-peaked profile of the [\ion{O}{iii}]$\lambda\lambda\,6300,6364$ doublet seen in the nebular spectrum of SN~2005bf~\cite{maedaetal2007}. This has also been recently observed in a nebular spectrum of an SLSN I, SN~2017gci~\cite{fioreetal2021}, but further polarimetric studies of the same object~\cite{pursiainenetal2023} disfavored this interpretation.}. Its data were interpreted with the synthesis of $0.6\,\msun$ of $^{56}\mathrm{Ni}$, but subsequently, Maeda~ et~al.~\cite{maedaetal2007} posited that the spin-down radiation of a newly born, highly magnetized NS (with a polar magnetic field of about $10^{14}$--$10^{15}\,\mathrm{G}$) was the major heating source of SN~2005bf. This scenario was also invoked for SN~2006aj, which was associated with an X-ray flash~\cite{mazzalietal2006}.

The magnetar model considers the spin-down radiation (Equation~(\ref{eq:sdlum1})) of a newly-born
and rapidly spinning magnetar as the chief heating source of the extreme SN. Despite what this description might suggest, pushing $B$ and $P$ to their extreme scales does not necessarily result in an SLSN. In fact, assuming that a core-collapse SN initially evolves as an adiabatic self-similar expansion at a velocity of $v_{\rm t}$, it will reach the maximum luminosity when all the entropy injected by the central power source (given by Equation~(\ref{eq:sdlum1}), in this case) is lost; this happens at a given radius ($R$) when the dynamical time scale ($t_{\rm m}$), i.e.,
\begin{equation}
    t_{\rm m}=\frac{R}{v_{\rm t}}\,,
\end{equation}
approximately equals the diffusion time scale of the photons ($t_{\gamma}$):
\begin{equation}
    t_\gamma=\left(\frac{\kappa M}{v_{\rm t}c}\right)^{1/2}\,,
\end{equation}
where $\kappa$ is the electron-scattering opacity, and $M$ and $E$ are the mass and kinetic energy of the ejecta, respectively. The magnetar looses its rotational energy (Equation~(\ref{eq:rotenergy})) on the spin-down time scale ($t_{\rm sd}$) (Equation~(\ref{eq:sdtime})). Hence, it results in 
the maximum luminosity~\cite{kasenandbildsten2010}:
\begin{equation}
    \label{eq:sdlum3}
    L_{\rm peak}=\frac{E_{\rm rot}t_{\rm sd}}{t_{\rm m}^2}\\\approx5\times10^{43}B_{14}^{-2}\kappa^{-1}_{0.2}M_5^{-3/2}E_{51}^{1/2}\,\mathrm{erg\,s^{-1}}\,.
\end{equation}

As can be seen in Equation~(\ref{eq:sdlum3}), values of $B\gg10^{14}\,\mathrm{G}$ make the SN fainter. This result is used later on in this work (see Section~\ref{sec:discussion}).

The magnetar model can explain a huge variety of light-curve shapes and evolutionary time scales, and for this reason, it is particularly flexible in accounting for the great diversity of SLSNe~I light curves (e.g.,~\cite{chatzopoulosetal2013,inserraetal2013,nicholletal2014,nicholletal2015,nichollandguillochonandberger2017}). However, it is difficult for the current knowledge to have strong independent constraints on the magnetar model for SLSNe. Theoretical efforts have shown possible independent signatures of a magnetar in the light curve~\cite{margalitetal2018,liuetal2021,gottliebandmetzger2024} (see also Section~\ref{sec:discussion}), the spectra~\cite{dessartetal2012,nicholletal2016,omandandjerkstrand2023} or outside the UV/optical/NIR wavelength range (e.g.,~\cite{margalitetal2018b}).
\subsection{Magnetar Scenario for LGRBs}
\label{sec:magnetargrb}
It is still an open question whether
the central engine of a GRB could harbor a rapidly spinning magnetar~\cite{usov1992,thompsonandduncan1993} rather than a rapidly accreting BH. The magnetar scenario could naturally explain some peculiar phenomenology observed in the X-ray emissions of GRBs. Indeed, the early evolution of long and short GRB afterglow ($<$ a few hours from the burst onset) observed in X-rays (0.1--10 keV) shows complex behavior that can be explained by the formation of a magnetar remnant, although alternative interpretations are plausible. Before the launch of the GRB dedicated space mission {\it Neil Gehrels Swift Observatory} ({\it Swift}~\cite{gehrelsetal2004})
, X-ray afterglows were observed only several hours after the GRB onset. With such a delay, afterglow radiation at a certain frequency ($\nu$) can be fairly described as $F(t,\nu)\propto t^{-\alpha} \nu^{-\beta}$ with $\alpha\sim1.2$--$1.5$ and $\beta\sim1$--$1.5$, in good agreement with synchrotron emission from an electron population energized in the interaction with the forward shock formed by the jet--ISM impact (e.g.,~\cite{Sari1998ApJ...497L..17S}). With the launch of {\it Swift} in November 2004, thanks to the fast slew capabilities of the payload, X-ray afterglows could be monitored with unprecedented timeliness, starting from a few minutes after the burst trigger, and revealed a complex behavior.  

The early X-ray emission of the vast majority of GRBs is described by a characteristic double broken power-law flux decay. Just after the prompt $\gamma$-ray emission, X-ray flux initially decays, following a very steep power law (much steeper than what is observed in late epochs), with $\alpha>2$ (sometimes $\alpha\gg2$) and with marked spectral evolution. After this ``steep'' decay phase and before the standard afterglow power-law decay phase observed after several hours, the majority of GRBs show a peculiar ``plateau''  ($0<\alpha<0.8$), which typically lasts for $\sim$$10^3$--$10^4$ s. Both the ``steep'' and ``plateau''\endnote{The so-called ``internal plateaus'' are shallow phases observed {\it before} the steep decay phase in a fraction of GRBs during the prompt phase, and their origin is sometimes attributed to energy injection from a magnetar (e.g.,~\cite{Gompertz2013MNRAS.431.1745G}). In these cases, if an afterglow plateau is present, it requires a different explanation.} phases challenge the standard afterglow interpretation~\cite{Nousek2006ApJ...642..389N,Gehrels2009ARA&A..47..567G}. While, today, there is a general consensus in interpreting the steep decay phase as high-latitude emission of the relativistic jet during the prompt phase 
(e.g.,~\cite{Lazzati2006ApJ...641..972L}), so far, no firm conclusion has been reached on the plateau's origin, and it is widely believed that it encodes crucial information on GRB physics.  

A shallow afterglow flux decay could, in principle, be obtained within the standard fireball model by assuming a wind environment and a very low-bulk Lorentz factor of the jet~\cite{Dereli2022NatCo..13.5611D}. In this scenario, the transition from the plateau to the normal afterglow evolution marks the crossing of a synchrotron-characteristic frequency that also generates a simultaneous spectral softening. However, one of the most intriguing properties of the majority of X-ray plateaus is that the spectral slope does not change when breaking to the normal decay phase, in marked contrast to the synchrotron expectation (e.g.,~\cite{Sari1998ApJ...497L..17S}). 
A different approach invokes a geometrical effect in the presence of a structured jet\endnote{A jet is called ``structured'' when the internal energy and expansion velocity gradually decrease with an increase in the angular distance from the jet axis, which is opposite to a ``top-hat'' jet, where the energy and velocity abruptly drop outside the jet cone.}. Indeed, for an observer slightly off-axis (i.e., with the line of sight slightly outside the jet core), an X-ray plateau could form in the afterglow evolution as an effect due to the continuous radiative supply from the innermost jet regions as soon as they decelerate enough to make the relativistic beaming comparable to the off-axis angle~\cite{beniaminiandramandeepandgranot2022}. These plateaus would be seen by a large fraction of observers and would last between $\sim$$100$ s and $\sim$10 ks. 

In a completely different paradigm, the X-ray plateaus could originate from a prolonged
 energy injection into the forward shock from a long-lived central engine. The source of energy could be a newly born magnetar 
\cite{usov1992,duncanandthompson1992,Kluzniak1998ApJ...505L.113K,Wheeler2000ApJ...537..810W}. The newly formed NS with millisecond spin periods is expected to loose its initial spin energy ($>$$10^{52}$ erg) at a very high rate for the first few hours through magnetic-dipole spin down, something that provides a long-lived central engine in a very natural way. Dai and Lu~\cite{Dai&Lu1998A&A...333L..87D} first considered this idea with regard to possible observable effects on the afterglow emission. Zhang and Meszaros~\cite{Zhang2001ApJ...552L..35Z} argued that, in this scenario, achromatic bumps in afterglow light curves are expected. Interestingly, studies of the origin of NS magnetism envisage that the millisecond spin period at birth is the key property that allows a proto-NS to amplify a seed magnetic field to a strength far exceeding $10^{14}$ G through efficient conversion of its initial differential rotation energy (e.g.,~\cite{duncanandthompson1992}). 
These highly magnetized, fast-spinning NSs are expected to lose angular momentum at a high rate in the first decades of their life and later become slowly rotating magnetars whose major free energy reservoir is in their magnetic field (see, e.g., \cite{mereghetti2008} and references therein). 
 
Analytical formulations of the injection of magnetar spin-down power into the afterglow emission component have been successfully tested over an extended sample of observed X-ray afterglow light-curve morphologies and luminosities (e.g.,~\cite{dallossoetal2011,rowlinsonetal2013,lietal2018,strattaetal2018,tangetal2019,ronchinietal2023}). Besides the good match between the analytical magnetar prescriptions on a large sample of observed X-ray afterglow light curves~\cite{dallossoetal2011,bernardinietal2012,strattaetal2018}, the inferred  magnetic field ($B$) and spin period ($P$) have also been found to be correlated in a way that matches the well-established ``spin-up line'' of radio pulsars and magnetic accreting NSs in galactic X-ray binaries (e.g.,~\cite{bhattacharayaandvandenheuvel1991,panandwangandzhang2013}) once re-scaled for the much larger mass accretion rates of GRBs~\cite{strattaetal2018}. 

The strong interactions between the magnetosphere and the accretion disk left behind by the progenitor are expected to produce a characteristic feature for an accreting NS. Indeed, a transition from accretion to a propeller regime, where disk material cannot enter the magnetosphere and the accretion power is reduced, is expected to produce a luminosity drop~\cite{illarionovandsunyaev1975,stellaandwhiteandrosner1986} that could, in principle, be tested during the prompt phase of GRBs. Bernardini~et~al.~\cite{bernardinietal2013} first demonstrated the general compatibility of this prediction by jointly considering both prompt and afterglow emissions for a small sample of long GRBs, where the prompt emission was assumed to be powered by accretion energy, while the afterglow plateau was assumed to be powered by the injection of the NS spin energy into the external shock
.  

This scenario was further elaborated and tested on a new sample of GRBs with evidence of an X-ray plateau~\cite{dallossoetal2023}. By assuming that the luminosity at the end of the prompt emission ($L_0$) is the luminosity at the transition to a propeller regime of an accreting magnetar, the analyzed GRBs were found to satisfy the so-called ``Universal'' relation in the plane:
\begin{equation}
\log L_0+\log P^{7/3}+\log R_6\quad\mathrm{vs}\quad\mu\,,
\end{equation}
where $R$ and $\mu$ are the NS radius and magnetic moment, respectively, which characterize different classes of known accreting sources in the propeller regime~\cite{campanaetal2018}. Through this relation, it was possible not only to verify the scenario in which accretion energy powers the prompt emission and the NS spin energy powers the afterglow plateau once accretion subsides but also to independently constrain the radiative efficiency of accretion in GRBs~\cite{dallossoetal2023}. 

Future multi-messenger observations of GRBs with plateaus, in synergy with third-generation gravitational wave detectors such as the Einstein Telescope~\cite{Punturo2010CQGra..27s4002P}, will help to shed light on this still debated phenomenology by simultaneously detecting continuous gravitational waves from a long-lived, fast-spinning NS remnant, if present. Among the most promising space mission projects for future GRB detection are THESEUS~\cite{Amati2018AdSpR..62..191A}, from which hundreds of GRBs per year are expected by the end of the 2030s, when third-generation gravitational-wave interferometers will be fully operative.
\section{GRBs and SNe: How Can a Single System Power Both? }
\label{sec:both}
The breakthrough in linking GRBs and SNe (e.g.,~\cite{woosleyandbloom2006, hjorthandbloom2012}) came with the association of GRB~980425 with SN~1998bw, a peculiar, energetic SN that exhibited an SN Ic BL spectrum and was co-located with the GRB~\cite{patatetal2001}, further supported by the analogous case of GRB~030329/SN~2003dh~\cite{valentitetal2008}. The GRB--SN connection is also supported by photometric and spectroscopic studies showing that the progenitors of both GRBs and their associated SNe exhibit aspherical features, with evidence of a conical structure in the bursts~\cite{Hoflichetal1999,Mazzalietal2005b}. Until now, observations of GRBs in association with SNe have primarily involved LGRBs and SNe Ic BL~\cite{woosleyandbloom2006}, which often surpass the energy of the relativistic jets by a factor of $10$--$100$~\cite{mazzalietal2014}. At the time of writing, the most preferred scenarios for these events are (i) the core-collapse (or BH-driven) model and (ii) the magnetar model. A key question is how the core collapse of a massive star can produce both am LGRB and an SN Ic BL. 
Both of them might offer reasonable pathways to power these extraordinary phenomena---in particular, during the same event.
 \subsection{Black Hole-Driven GRB SNe}
 \label{sec:bhdriven}
 In the case of a core-collapse explosion, a mechanism that can potentially account for both phenomena is the formation of an accretion disk around a BH during the collapse. The disc, indeed, would be able to power both the relativistic jet and the SN.  However, the creation of the disk strongly depends on the angular momentum of the star. Woosley and Heger~\cite{woosleyandheger2006} showed that a star must have a rotational period of $\sim$$1$~ms to provide an explosion energy of $\sim$$10^{52}$~erg, as was inferred for some of the SNe accompanying GRBs. This energy is likely to originate from viscous heating within the accretion disk~\mbox{\cite{macfadyenandwoosley1999,pophamandwoosleyandfryer1999, kohriandnarayanandpiran2005, Menegazzietal2024a, Menegazzietal2024b}.}

In this model, a GRB accompanied by an energetic SN has three components. The first is a narrow, ultra-relativistic central jet with a Lorentz factor of $\gtrsim$$300$ and an opening angle of $\sim$$0.1$ radians around the rotational axis, responsible for the LGRB and carrying a small mass fraction (less than $10^{-6}\,\mathrm{M}_\odot$). 
{{The second is a broader region of extremely energetic (with kinetic energy of $\sim$$10^{52}$~erg) ejecta that extends out the rotational axis to $\sim$$1$~radians 
and lies just outside the innermost region of the accretion disk (extending up to several thousand kilometers from the BH), where viscous heating, neutrino-driven winds and shock interactions lead to the ejection of material. This region}} carries a significant portion of the progenitor mass (approximately $10\,\mathrm{M}_\odot$) at sub-relativistic speeds (around \mbox{10,000--20,000$\,\mathrm{km,s^{-1}}$}), and it is responsible for the SN emission and $^{56}$Ni production~{\cite{SurmanandMcLaughlin2005, Surmanetal2006, SurmanandMcLaughlinandSabbatino2011, SongandLiu2019}. {The production of $^{56}$Ni may occur either in the hot disk outflows}\endnote{{{The interested reader can refer to the simulations performed by Menegazzi}~\cite{Menegazzietal2024a} {to see the $^{56}\mathrm{Ni}$ production in the outflow launched by a BH+disc engine.}}} {or in shock-heated stellar material near the base of the jet, and it has been shown by both simulations and semi-analytical studies to strongly depend on the disk accretion rate}~\cite{Surmanetal2006, SurmanandMcLaughlinandSabbatino2011, SongandLiu2019}.} Finally, the third region is the non-relativistic ejecta along the equatorial plane naturally present in models where the outflow at the equator is blocked~\cite{woosleyandbloom2006}. This dual-outflow nature of the explosion---where one outflow produces the SN with massive sub-relativistic ejecta and the other forms a relativistic jet that generates the GRB---raises significant questions regarding the processes driving both phenomena. While it is clear that these compact objects are necessary for generating relativistic jets, the energy sources driving both the SN and the jet are still not fully understood. One of the main reasons is that the large amount of energy released by the SN exceeds the amount predicted by the commonly accepted neutrino-driven explosion mechanism. A possible alternative considers that the jet substantially contributes to the SN explosion~\cite{leblancandwilson1970, khokhlovetal1999, burrowsetal2007, gilkisandsoker2014}. Some numerical simulations, where the jet is manually introduced at the core of the star, have been performed, and they have shown that a collimated jet, if it successfully breaks out through the stellar envelope, can deposit enough energy to unbind the entire star, leading to the formation of both a GRB and an SN explosion~\cite{lazzatietal2012}. This process has been observed not only in regular LGRBs but also in choked jet events such as LLGRBs. More studies on the dynamics of SNe driven by relativistic jets have been carried out by Barnes~et~al.~\cite{barnesetal2018} and Shankar~et~al.~\cite{shankaretal2021}, who focused on the light curves and spectra of SNe Ic BL. They found that a single central engine, likely a rapidly spinning NS or BH, can produce both a GRB and an SN Ic BL by powering an ultra-relativistic jet. This jet can simultaneously explode the progenitor star, generating an SN with features typical of SNe Ic BL, and driving a LGRB. Finally, it remains unclear whether the jet and the SN are produced simultaneously or whether there is a temporal separation between the two outflows~\cite{hjorth2013} and whether there is any interplay between the two components~\cite{decolleetal2022}.

If this scenario is valid, it is then licit to ask why only a small fraction of massive stars are seen as GRB SNe. A first answer could be given by the environmental metallicity. Besides the rotation, the environmental metallicity ($Z$) also plays a crucial role in determining whether a collapsing star can generate a GRB with an SN. Low-metallicity (sub-solar) environments promote the formation of rapidly rotating massive stars by reducing stellar winds, which are primarily driven by metal-line opacities. Therefore, stars in low-metallicity host galaxies lose less mass to stellar winds compared to those located in regions with higher metallicity, being more likely to collapse directly into a BH with high angular momentum~\cite{hegeretal2003, Eldridgeandtout2004, hjorthandbloom2012, Pejchaandthompson2015}.

On the contrary, high-metallicity stars are more likely to lose angular momentum through wind-driven mass loss, preventing the formation of relativistic jets and favoring a more spherical, lower-energy SN explosion without a GRB. Woosley and Bloom and Hjorth and Bloom~\cite{woosleyandbloom2006, hjorthandbloom2012} showed that GRB SNe typically occur in host galaxies whose metallicity is about $0.2$--$0.4\,{\rm Z_\odot}$ (where ${\rm Z_\odot}\approx 0.02$ is the metallicity of the Sun~\cite{Vagnozzi2019}). In addition, the efficiency of $^{56}$Ni nucleosynthesis can also affect the visibility of the SN signal. Recent studies have shown that the wind generated by BH-disk systems in failed SNe, which is thought to be the origin of the jets in the collapsar model, is rich in $^{56}$Ni ($\ge$$0.1\, {M}_\odot$)~\cite{hayakawaandmaeda2018, justetal2022, fujibayashietal2024, deanandfernandez2024}. This suggests that measuring the amount of $^{56}$Ni produced during an explosion can provide significant constraints on the connection between GRBs and SNe. Tominaga~et~al.~\cite{tominagaetal2007, tominaga2008} studied the jet-induced explosions of a Population III $40\,\mathrm{M}_\odot$ star. Their results suggested a correlation between GRBs with and without bright SNe and the energy deposition rate ($\dot{E}_\mathrm{dep}$; see also~\cite{maedaandnomoto2003, nagatakietal2006}). They found that the energy deposition rate significantly influences the synthesis of $^{56}$Ni.
In explosions with high $\dot{E}_\mathrm{dep}$ ($\dot{E}_\mathrm{dep}\gtrsim 6\times 10^{52}$ erg), a large amount of $^{56}$Ni ($\gtrsim$$0.1\,\mathrm{M}_\odot$) was synthesized, consistent with GRB--SNe observations.  Contrarily, in explosions with lower $\dot{E}_\mathrm{dep}$  ($\dot{E}_\mathrm{dep}\lesssim 3\times 10^{51}$ erg), the $^{56}$Ni mass is much smaller ($\lesssim$$10^{-3}\,\mathrm{M}_\odot$), comparable with observations of GRBs that do not exhibit bright SN signatures, such as GRB~060505 and GRB~060614.
\subsection{Magnetar-Driven GRB SNe}
\label{sec:magdriven}
Magnetars have been found to be a compelling explanation for the central engine driving both GRBs and their associated SNe, particularly in the context of SNe Ic BL~\cite{bucciantinietal2009, woosley2010, metzgeretal2011}.
The rotational energy (Equation~(\ref{eq:rotenergy})) of a millisecond magnetar can represent a natural source GRB energy ($\sim$$10^{51}$--$10^{52}$ erg) and SNe ($\sim$$10^{51}$ erg)~\cite{metzgeretal2011, obergaulingerandaloy2020}, and their rapid spin periods and high magnetic fields generate strong magnetic dipole radiation and Poynting flux-dominated jets capable of accelerating particles to relativistic speeds and forming the observed GRB outflows. In particular, the formation of relativistic jets in the magnetar scenario is facilitated by magneto-centrifugal processes and magnetic reconnection, which create a structured jet that can naturally explain the observed variability and polarization in GRBs~\cite{metzgeretal2011, Mundelletal2013}. Unlike BH-driven models, which rely on accretion, magnetars derive energy from their spin-down; hence, they are able to sustain the injection of energy into the jet over time scales corresponding to their spin-down period, producing extended emission phases seen in GRB afterglows (see Section~\ref{sec:magnetargrb}).
Recent hydrodynamic simulations (e.g.,~\cite{bucciantinietal2009, metzgeretal2011}) have demonstrated that the energy released by a magnetar can produce a sufficiently powerful shock that accelerates the surrounding material to relativistic speeds. Magnetar-driven jets can penetrate the progenitor star’s envelope, and this interaction results in so-called ``magnetar shock breakout'', producing the observed high-energy $\gamma$-ray emission while simultaneously depositing energy into the SN ejecta. The energy budget of these jets, estimated to be $\sim$$10^{51}$--$10^{52}$~erg, is consistent with the observed luminosities of LGRBs~\cite{thompsonandchangandquataert2004, margalitetal2018, metzgeretal2011}. The combination of the relativistic jet and the energetic SN explosion can account for the diverse range of GRB--SN observations, including the different luminosities, spectral features and light-curve shapes observed across various events (e.g.,~\cite{bucciantinietal2009}).

Several observations support the idea that magnetars are, indeed, the central engines of GRB SNe. A first piece of evidence comes from a detailed study of the SN associated with GRB~130427A, as its high luminosity and broad spectral lines are possibly indicative of the involvement of a rapidly spinning magnetar (e.g.,~\cite{mazzalietal2014, Bernardinietal2014}). The rapid rise and decline in the light curve further align with the predicted energy injected from a magnetar, which releases energy into the surrounding material over time as it spins down due to magnetic braking. Furthermore, the study of SN~1998bw, the SN associated with GRB~980425, further supports the magnetar model, as it successfully explains its high-energy output (e.g.,~\cite{Wheeler2000ApJ...537..810W}). The unusually broad-lined spectrum and high kinetic energy of the ejecta of SNe like that associated with GRB~130427A and SN~1998bw are difficult to reconcile with standard collapsar models, but they instead foster the magnetar scenario as the mechanism powering these explosions. The idea that a magnetar can power both the GRB and the associated SN was further reinforced by other observations. In the case of ULGRB GRB~111209A, the spectral signatures and the temporal evolution of the X-ray afterglow, along with the measure of environmental density, suggest ongoing energy injection by a highly magnetized, rapidly spinning NS (e.g.,~\cite{gaoandyouandwei2016, GompertzandFruchter2017}). Additionally, Zhang~et~al.~\cite{Zhangetal2022} showed that the optical--UV light curves of SN~2006aj and the afterglow emission of the correlated GRB~060218 can be explained by a magnetar scenario. More examples corroborating this hypothesis were provided by Kumar~et~al.~\cite{Kumaretal2022} in analyzing the prompt characteristics and the late-time optical follow-up observations of GRB 171010A/SN~2017htp and GRB 171205A/SN~2017iuk. 

Additional clues pointing at magnetar-powered GRB SNe come from the diversity of their usually bright light curves. An example is event SN~2011kl (associated with GRB~111209A; see Section~\ref{sec:11kl}). The afterglow and thermal emission peaks in SN~2011kl are consistent with energy input from a magnetar engine rather than the fallback accretion model traditionally invoked in GRBs. Theoretical work has shown that the luminosity and kinetic energy of GRB SNe can be naturally explained by the energy released through the decay of magnetic fields and rotational energy in magnetars. This energy, estimated to be of the order of $10^{51}$--$10^{52}$~erg, can be efficiently channeled into the surrounding ejecta, leading to the rapid acceleration and eventual explosion of the progenitor star (e.g.,~\cite{kasenandbildsten2010}). Observations of GRB~161219B/SN~2016jca provide further support, as the detailed modeling of its light curve required a long-lasting energy injection consistent with the magnetar mechanism~\cite{Ashalletall2019}.
In addition to the X-ray plateau phase and the SN light curves, the spin-down activity, along with the continued fallback accretion that takes place in the magnetar scenario, can reproduce unexpected increases in brightness occurring hours to days after the initial burst observed in some GRB light curves, known as ``rebrightenings''~\cite{RamirezRuietal2001, Lazzatietal2001}. Observations such as the late-time optical rebrightening in GRB~100814A, which was interpreted as the result of a spin-up magnetar due to fallback accretion, further support this theory. GRB~161219B/SN~2016jca, which displayed a similar late-time optical rebrightening, has also been modeled with fallback accretion onto a newly born magnetar~\cite{Yuetal2015, Ashalletall2019}. Another feature observed in GRBs that supports the magnetar scenario is the detection of polarized $\gamma$-ray and X-ray prompt and afterglow emissions in GRBs. Such signatures, indeed, suggests the presence of ordered magnetic fields, consistent with the magnetar-driven jet model. For instance, high polarization fractions in GRB~020813~\cite{bjornsson2003,covinoetal2003} and GRB~021206~\cite{Coburnandboggs2003, boggsandcoburn2003, Rutledgeandfox2004}, although typically associated with limited statistical significance \citep{McConnell2017}, may indicate strong magnetic-field alignment in relativistic jets~\cite{Kumarandpanaitescu2003,Zhangandmeszaros2004}. The first polarimetric observations of an X-ray afterglow were obtained with the Imaging X-ray Polarimetry Explorer (IXPE) for GRB 221009A, for which an upper limit on the polarization degree of $13.8\%$ was provided in the $2$--$8$ keV energy range~\cite{Negro2023}.
\section{Review of Some Peculiar GRB--SN Associations}
\label{sec:pecgrbsne}
\subsection{GRB 101225A: The Christmas Burst}
On 25 December 2010 at the time $T_0 = 18$:37:45 UT~\cite{racusinetal2010} the BAT telescope triggered Swift to observe powerful burst GRB~101225A; besides its incidental discovery on Christmas (hence, often referred to as ``the Christmas burst''), GRB 101225A owes its fame to its exceptionally long duration, with $T_{90}>2000\,\mathrm{s}$~\cite{thoeneetal2011b}. GRB~101225A also showed a peculiar $\gamma$-ray light curve with a plateau extending up to $\sim$$1700\,\mathrm{s}$, after which the X-ray light curve showed flaring, with a flux decaying in time as $\sim$$t^{-1}$~\cite{levanetal2014}; interestingly, the X-ray light curve closely resembles that of GRB~111209A (see Section~\ref{sec:11kl}). Soon after $T_0$, its emission was detected at longer wavelengths and could be observed up to approximately two months after $T_0$. The light curve of its optical afterglow showed a flattening about 10 days after $T_0$ coinciding with the appearance of a new additional component in the spectral energy distribution (SED). As the redshift of GRB 101225A could not be measured from the host spectrum (see later), Th\"one~et~al.~\cite{thoeneetal2011b} estimated it through the comparison of the light curve and the SED with those of GRB 980425/SN~1998bw (see Supplementary Information in~\cite{thoeneetal2011b}) and obtained $z=0.33$. This resulted in an absolute magnitude of the associated SN at a peak of $M_B\simeq-16.7\,\mathrm{mag}$, making it the faintest GRB SN observed at that time. Furthermore, Th\"one~et~al. put the GRB 1012225A SN in comparison with other GRB~SNe using the $(s,k)$ formalism introduced by Zeh, Klose and Hartmann~\cite{zehandkloseandhartmann2004}\endnote{In the $(s,k)$ formalism, $s$ and $k$ are time-stretching and luminosity-scaling factors, respectively, which fit the light curve of SN~1998bw to that of a given SN. Hence, by definition, SN~1998bw has $s=k=1$.} and found $s=1.25$ and $k=0.08$ for the GRB 101225A SN, corresponding to a luminosity about 12.5 times fainter than that of SN~1998bw.

\begin{figure}[H]
    
    \includegraphics[width=0.8\linewidth]{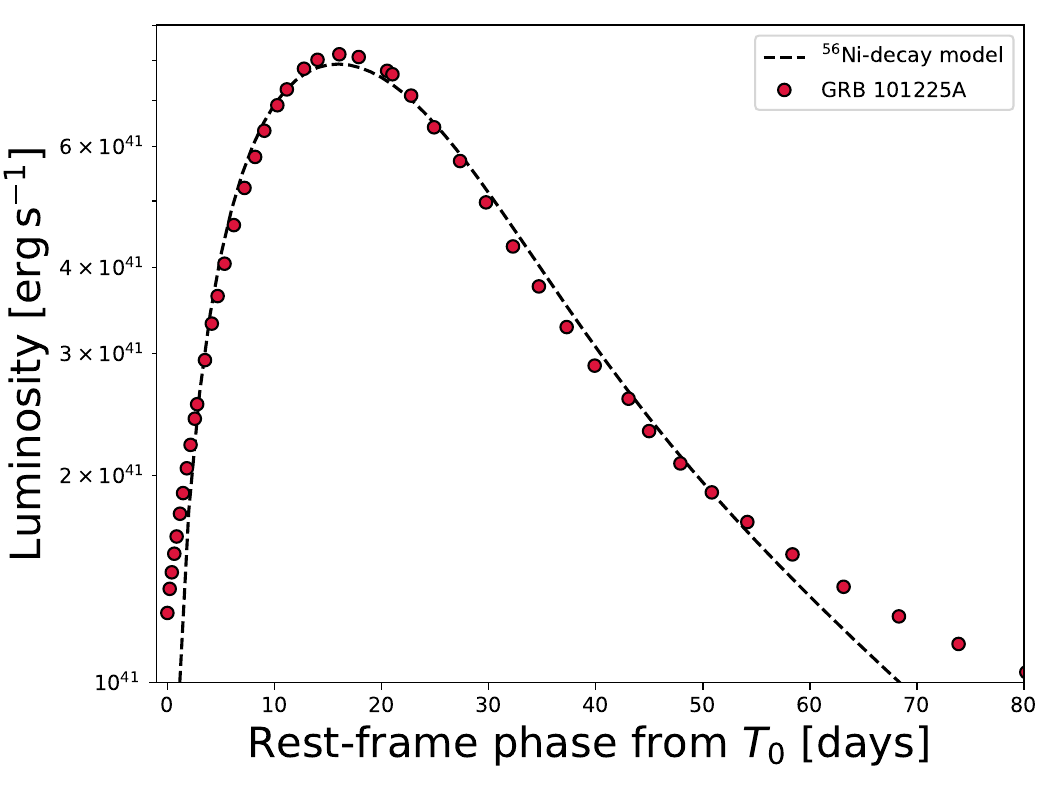}
    \caption{The bolometric ($UBgriz$) light curve of the GRB~101225A SN fitted with a $^{56}$Ni decay-powered diffusion scheme (dashed black lines). The original multiband light curves were taken from Th\"one~et~al.~\cite{thoeneetal2011b} and integrated to obtain the bolometric light curve (red dots) assuming a distance of 1.6 Gpc, then corrected for time delay. }
    \label{fig:grb101225asn}
\end{figure}

\textls[-25]{Moreover, a spectrum of GRB 101225A was observed on $\sim$$T_0+51$ h with GTC+OSIRIS~\cite{cepaetal2000} and showed no unambiguous lines/features but an optical blue continuum.} The SED of GRB 101225A is well described by an absorbed power law and a black body in the X-rays with a cooling and expanding black body between UV and NIR bands until the additional component shows up~\cite{thoeneetal2011b}. While different explanations have also been put forward~\cite{campanaetal2011}, Th\"one~et~al. suggested that both the X-ray and the UVOIR thermal components of GRB 101225A can be seen as a natural consequence of the merging of a compact object with a helium star. To reach such a configuration, two massive stars orbiting each other remain bound after the SN explosion of one of them, creating a new binary system made of a degenerate and a massive star. Upon the latter star’s departure from the main sequence, the system enters a common envelope phase in which the compact remnant gradually reaches the center and, due to its angular momentum, accretes matter via a disk~\cite{fryerandwoosley1998}. This configuration may be suitable to reproduce the observed data of GRB 101225A, since (i) it may allow for the launch of a relativistic jet that is, in principle, able to power a GRB; (ii) if the remnant is a magnetar, its spin-down radiation may boost the prolonged emission (see Section~\ref{sec:magnetargrb}); (iii) it naturally explains the presence of the two thermal components as the result of the interaction between the relativistic jet and the common-envelope material ejected by the progenitor; and (iv) detailed calculations by Barkov \and Komissarov for a helium star--BH merger~\cite{barkovandkomissarov2011} show that the amount of $^{56}$Ni synthesized by these phenomena is limited to a few times $0.02\,\msun$, in agreement with a faint GRB--SN companion. This is consistent with our estimate of the $^{56}$Ni mass needed to power the GRB 101225A SN ($M_{56{\rm Ni}}=0.036\,\msun$) which we obtained by fitting its bolometric light curve obtained by Th\"one~et~al. with the TigerFit tool~\cite{chatzopoulosetal2013} (see Figure~\ref{fig:grb101225asn}).

The compact--He star merger scenario is not the only one proposed to explain the Christmas burst: L\"u~et~al.~\cite{lueetal2018} suggested that GRB~101225A could be powered by a nascent magnetar with $B<5.8\times10^{15}\,\mathrm{G}$ and $P_0<1.25\,\mathrm{ms}$ collapsing into a BH (see also~\cite{zouetal2021}). As already discussed in Section~\ref{sec:magnetarsn}, in the general case, a newly born magnetar looses energy via both gravitational and magnetic-dipole radiation, with different weights that vanish and dominate as time goes by. In the two limiting cases in which the energy losses are purely gravitational or radiative, the X-ray luminosity declines with a slope $\alpha=-1$ or $\alpha=-2$, respectively. Should the magnetar collapse into a BH before entering the radiatively dominated phase, the $\alpha=-2$ phase is absent, and, due to the sudden quenching of the radiative losses, an abrupt steepening of the afterglow light curve should be visible (see Figures~1 and 2 in~\cite{lueetal2018}). 
\subsection{GRB~111209A/SN~2011kl}
\label{sec:11kl}
On 9 December 2011 at time $T_0 = 07$:12:08 UT, the Swift satellite~\cite{gehrelsetal2004} equipped with the Burst Alert Telescope (BAT)~\cite{barthelemyetal2004} discovered unusually long-lasting GRB 111209A~\cite{hoverstenetal2011,palmeretal2011} at a redshift of $z=0.677$~\cite{vreeswijketal2011}. It was actually observed about $\sim$$5400\,\mathrm{s}$ earlier~\cite{golenetskiietal2011} in $\gamma$ rays by the Konus Wind spectrometer~\cite{aptekaretal1995}, corresponding to an overall duration of the prompt emission of about 4 h in the rest frame~\cite{gendreetal2013}. The host galaxy of GRB 111209A is compact and faint ($M_B\approx-17.6$ mag), with a moderately sub-solar metallicity of $Z\approx0.35\,Z_\odot$ at the GRB location~\cite{strattaetal2013,levanetal2014}. The $\gamma$-ray light curve presented a double peaked profile with a prominent flare at $\sim$$T_0+2\,\mathrm{ks}$ first detected in Konus Wind data~\cite{golenetskiietal2011}, then in the optical bands with a phase lag of $\sim$$0.410\,\mathrm{ks}$~\cite{strattaetal2013}. Gendre~et~al.~\cite{gendreetal2013} discussed the X-ray observations of GRB 111209A obtained with the Swift/X-ray telescope (XRT) in detail~\cite{burrowsetal2005} via the XMM Newton satellite archive, as well as optical data obtained with TAROT~\cite{klotzetal2009} and Swift/UVOT, to which a number of optical measurements previously published by Klotz~et~al.~\cite{klotzetal2011} and Kann~et~al.~\cite{kannetal2011} were also added. The X-ray data presented by Gendre~et~al. sample the emission of GRB 111209A up to $\sim$$T_0+26$ days; throughout its evolution, the light curve approximately follows different power laws ($\propto t^{-\alpha(t)}$), similar to many other GRBs~\cite{liangetal2010}. In the case of GRB 111209A, the light curve initially follows a shallow decay with an index of $\alpha(t)\approx0.544$. Then, at about $T_0+5.5$ h\endnote{Gendre~et~al. computed the duration of the prompt phase up to this point, corresponding to $T_0\,+$~20,000$\,\mathrm{s}$.}, the decaying light curve steepens ($\alpha(t)\approx4.9$) and smoothly settles on a plateau ($\alpha(t)\approx0.5$) before entering the afterglow phase ($\alpha(t)\approx1.5$)~\cite{gendreetal2011}. Such behavior suggests the presence of multiple mechanisms operating on different time scales, with one superseding another. For instance, the sudden steepening of the light curve after the prompt phase is usually attributed to a high-latitude emission from a relativistic jet 
coming from a direction of $\gg$$\Gamma^{-1}$~\cite{kumarandpanaitescu2000}). The spectrum of GRB 111209A is well described by the combination of a broken power law and a thermal blackbody component between $T_0+425\,\mathrm{s}$ and $T_0+1000\,\mathrm{s}$. However, it marginally contributes (about 0.01\%) to the total emission between $0.5$ and $10\,\mathrm{keV}$ and vanishes in later epochs. Furthermore, the presence of a thermal component was also excluded in the optical bands after having extracted the SED via optical photometry.

\begin{figure}[H]
    
    \includegraphics[width=0.7\linewidth]{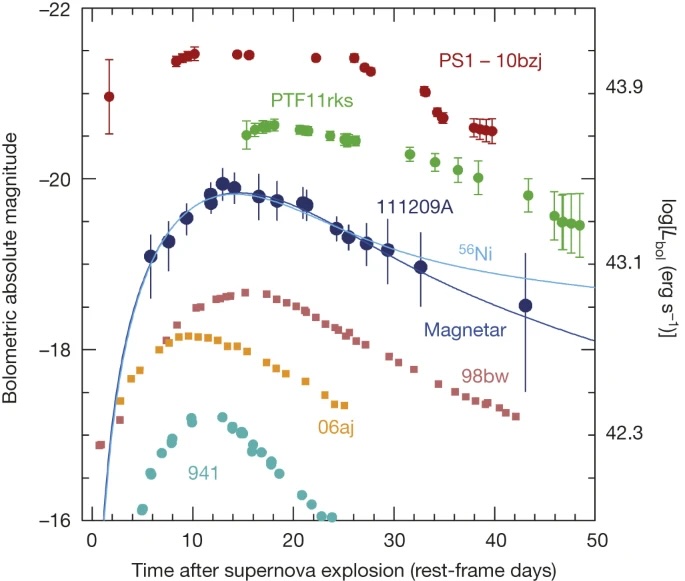}
    \caption{The bolometric light curve of GRB~111209A/SN~2011kl (blue circles) compared with those of GRB 980425/SN~1998bw (red squares) and of soft GRB (X-ray flash) XRF 060218/SN~2006aj (yellow squares), as well with GRB-less supernovae as SN Ic 1994I (turquoise circles), SLSNe I PTF11rks (green circles) and PS1-10bzj (red circles). All light curves are  integrated over the same wavelength bands with 1$\sigma$ error bars. For SN~2011kl, the best-fitting models assuming either a magnetar (blue line) or $^{56}$Ni (light blue line) power sources are also shown. This figure from~\cite{greineretal2015} (Figure 2 in the original paper) was reproduced with permission from Springer Nature.}
    \label{fig:2011kl}
\end{figure}

Additional data of GRB 11209A from Levan~et~al.~\cite{levanetal2014} were also analyzed by S\mbox{tratta~et~al.~\cite{strattaetal2013}}. In these preliminary dataset, the photometric points were densely sampled up to the early afterglow phases (at $\sim$$T_0+17$ h) and thinned out thereafter. Altogether, the data presented by Gendre~et~al. and Stratta~et~al. encompass a significant portion of the evolution of GRB 111209A, but they did not provide definitive evidence of an associated SN. About one year later, Greiner~et~al.~\cite{greineretal2015} presented unpublished optical/near-infrared photometry taken with the seven-channel GROND imager~\cite{greineretal2008} and a UV/optical VLT+X-Shooter~\cite{vernetetal2011} spectrum observed on 29 December 2011 (corresponding to $T_0+11.8$~days). In these new data, it was clear how the afterglow light curve deviated from the power-law decay at about $T_0+15$ days while exhibiting a hump similar to that observed in the case of GRB~SNe (see Figure~\ref{fig:pozanenko}). The association of an SN with a ULGRB was as unprecedented, as it was the association of an SLSN I with a GRB. In fact, after having disentangled the afterglow emission, the light curve of the companion SN, SN~2011kl, reached the unusually bright bolometric peak of approximately $-20\,\mathrm{mag}$, which has never been observed in hypernovae/SNe Ic BL and is more typical of SLSNe (see Section~\ref{sec:slsne} and Figure~\ref{fig:2011kl}). In addition, the VLT+X--Shooter spectrum of SN~2011kl, corresponding approximately to two days before its maximum luminosity, is much bluer than those of ordinary hypernovae; both these photometric and spectroscopic features are more similar to those of SLSNe I compared to standard hypernovae/SNe Ic BL. This was also confirmed by the analysis of Liu~et~al.~\cite{liuetal2017}, who compared the spectrum of SN~2011kl with average spectra of a sample of SNe Ic BL and SLSNe I. This spectrum does not unambiguously show the prominent \ion{O}{{ii}
} absorptions usually seen on the blue side of SLSNe I optical spectra {around maximum}, but given the poor signal-to-noise ratio of the spectrum, it is difficult to attempt a careful line~identification.
The biggest difficulty in the interpretation of GRB 111209A is accounting for such a long-lasting power source. Various models have been proposed to account for both SN~2011kl and the counterpart of GRB~111209A. Before the association with an SN became clear
, Gendre~et~al. and Stratta~et~al. had posited that a low-metallicity blue supergiant star~\cite{woosleyandheder2012} might be the progenitor of GRB 111209A. This scenario was the most plausible by exclusion compared to other possible mechanisms that could provide long-lasting power sources, such as a a pair-instability SN explosion (see Section~\ref{sec:slsne}) or supergiant/helium star systems in tidally locked binaries~\cite{gendreetal2013}. Among these scenarios, the pair-instability explosion was initially disfavored given the early absence of an SN companion. Furthermore, the absence of a prominent thermal component\endnote{Even though a thermal component is visible throughout the spectral evolution of GRB 111209A, it is subdominant and disappears in the high-energy bands after 5.5 h.} in the SED and the faintness of the host galaxy of GRB 111209A do not support these scenarios~\cite{starlingetal2012}. 

The interpretation of the observed data with the $^{56}\mathrm{Ni}$ decay model led to a $^{56}\mathrm{Ni}$ mass of $\sim$$1.0\pm0.1\,\msun$, but this interpretation was disfavored based on the observed UV spectra, which have a lower degree of suppression compared to the standard, SN~1998bw-like hypernovae and, pointing at a low metal content. Therefore, Greiner~et~al. suggested that the maximum luminosity of SN~2011kl was powered by the spin-down radiation of a millisecond magnetar (see also~\cite{kasenandbildsten2010,woosley2010}). However, Ioka~et~al.~\cite{Iokaetal2016} showed that this magnetar interpretation presents a significant challenge: the spin-down time required to power SN~2011kl is $\sim$$1.1\times10^6$~s (about 13 days), whereas GRB~111209A lasted only $\sim$$10^4$~s. This two-orders-of-magnitude discrepancy raises questions about whether a single magnetar could be responsible for both the GRB and the SN-like bump. If the same magnetar were to explain both, it would require an initial rapid release of energy to power the GRB, followed by a much slower energy injection to sustain the SN-like emission, implying an unusual and fine-tuned evolution of its magnetic field. In the literature, different authors have considered the contribution of $^{56}\mathrm{Ni}$ decay in SN~2011kl either as the primary heating source~\cite{greineretal2015,kannetal2018,wangetal2017} and found $^{56}\mathrm{Ni}$ best-fit masses of approximately $1$--$5\,\msun$ or a secondary contributor subordinate to the heating of a millisecond magnetar~\cite{metzgeretal2015,berstenetal2016,wangetal2017}, with a fixed mass of about $0.1$--$0.2\,\msun$. Differences in the estimates of the $^{56}\mathrm{Ni}$ mass are due to, e.g., the use of different setups, different assumptions on the opacity or even different bands over which the photometry was integrated to obtain the bolometric light curve.

Alternatively, the emission of GRB 111209A could be interpreted with the inclusion of an additional component usually absent in normal LGRBs, like a SN shock breakout~\cite{campanaetal2006,starlingetal2011}, a TDE~\cite{burrowsetal2011,bloometal2011,zaudereretal2011,levanetal2011,thekhovskoyetal2014} or fallback accretion onto a BH left as aftermath of the core collapse of $\sim$$10\,\mathrm{M_\odot}$ Wolf--Rayet (WR)-like progenitor. In the latter case, the ULGRB is modeled as a normal collapsar in which part of the unbound material falls back and accretes onto the BH, potentially allowing for the prolonged emission of GRB~111209A.

More recently, however, studies such as that of Moriya~et~al.~\cite{Moriyaetal2020} have reconsidered the PPI mechanism as a viable explanation for both the long duration of GRB~111209A and the exceptionally bright nature of SN~2011kl. Indeed, they suggested that, rather than a pair-instability SN fully disrupting the progenitor, a PPI event could have led to the formation of a massive extended envelope, significantly altering the final core-collapse dynamics. In their analysis, Moriya~et~al. showed that some rapidly rotating, hydrogen-free gamma-ray burst progenitors can experience PPI shortly before core collapse. These progenitors can maintain an extended structure up to thousands of solar radii, making them ideal candidates for ULGRBs. Specifically, in their simulations, a helium star with an initial mass of 82.5 $\mathrm{M}_\odot$ evolved into a 50 $\mathrm{M}_\odot$ progenitor with an extended envelope of thousands of solar radii (reaching 1962 $\mathrm{R}_\odot$) at collapse. When the explosion occurs, the shock breakout leads to a long-lasting cooling phase, resulting in a rapidly evolving ($\lesssim$10 days) and luminous ($\gtrsim$$10^{43}$ erg s$^{-1}$) optical transient. This extended cooling phase post explosion can account for the rapid rise and bright peak of SN~2011kl, without requiring an extreme $^{56}$Ni mass. As an alternative model, they also considered the collapsar scenario, in which the ultra long duration of GRB~111209A would be attributed to sustained fallback accretion. However, Moriya~et~al. argued that the extended hydrogen-free progenitor model provides a more self-consistent explanation, as it can simultaneously explain the long-lived GRB emission and the fast-evolving, luminous SN counterpart. As a matter of fact, their model successfully reproduces the rapid rise and bright peak of SN~2011kl and accounts for its slow decline when energy input from $^{56}$Ni decay is considered. Furthermore, their work suggests that when the GRB jet is choked or viewed off-axis, the resulting event could manifest as a fast blue optical transient (e.g.,~\cite{Lyutikov2022}) rather than a classical GRB--SN association. This mechanism naturally explains why not all ULGRBs have associated SNe and why some SNe appear particularly luminous, even in the absence of strong GRB signals. By integrating these aspects, their findings reinforce the idea that GRB~111209A/SN~2011kl originated from a helium star with a highly extended envelope, supporting an alternative formation channel distinct from standard blue supergiant and WR progenitors. 
\subsection{GRB 140506A}
\textls[-30]{GRB 140506A triggered the Swift satellite on 6 May 2014~\cite{gompertzetal2014,markwardtetal2014} at $T_0=21$:07:36 {UT}. Its host galaxy is metal-poor ($Z\approx 0.35\,Z_{\rm \odot}$) and moderate star-forming ($\mathrm{SFR}\approx1.34\,\mathrm{M_\odot\,yr^{-1}}$), making it unexceptional among the LGRB host galaxy population~\cite{heintzetal2017}. Its prompt emission was also detected by the Konus Wind and Fermi~\cite{golenetskiietal2014,jenke2014} satellites and lasted $T_{90}=111$ s; overall, its duration, SED and isotropic energy release~\cite{kannetal2024} in red/NIR bands mirror those typical of LGRBs, but its spectrum exhibited quite unusual features for this subclass. Fynbo~ et~al.~\cite{fynboetal2014} presented and discussed photometric data of GRB 140506A observed with GROND in the $g',r',i',z',J,H$ and $K_{\rm s}$ filters and an additional unfiltered one with the IMACS instrument mounted on the Magellan--Baade telescope at Las Campanas Observatory~\cite{dressleretal2011}. These data are extended up to 68 days after $T_0$. They also obtained two X-Shooter spectra with a very high signal-to-noise ratio at $T_0+8.8$ h and \mbox{$T_0+33$ h} over the entire UV-NIR (300--2500 nm) range, as well as a lower-quality one with the Magellan telescope at $T_0+52$ days. The former were also used to measure the redshift ($z=0.88911$) based on the [\ion{O}{ii}] $\,\lambda\lambda\,3727,3729$ emission doublet from the host galaxy and some absorptions attributed to the afterglow. Among these, the identification of Balmer and excited \ion{He}{I} absorption lines was unprecedented for the spectrum of a GRB afterglow. Furthermore, GRB 140506A exhibited a quite unusual SED with a strong UV suppression, which could not be described by any extinction model for the local group~\cite{heintzetal2017}. However, the authors attributed these peculiarities to line-of-sight effects\endnote{Based on their analysis, Fynbo~et~al.~\cite{fynboetal2014} ascribed the Balmer and excited \ion{He}{i} absorption lines to the presence of a \ion{H}{ii} region and an associated partially ionized zone/photodissociation
zone, respectively, whereas the UV-flux suppression was due to the absorption from a cooler region (although this last point was disfavored by Heintz~et~al.~\cite{heintzetal2017}).}.}

\textls[-15]{Approximately one year after $T_0$, further observations of the same field were conducted to study the host galaxy of GRB 140506A. Heintz~et~al.~\cite{heintzetal2017} noticed that the magnitudes of the optical host galaxies were about one magnitude fainter than the previous measurements presented by Fynbo~et~al. and tentatively suggested the possibility of a bright GRB~SN accompanying the GRB $\sim$$35$ rest-frame days after $T_0$. These findings were fostered by the analysis of Kann~et~al.~\cite{kannetal2024}, who presented new Swift/UVOT dataset starting from $T_0+108$ s (see also
\cite{siegelandgompertz2014}) and re-analyzed the previously published X-ray/UV/optical/NIR photometry of GRB 140506A~\cite{fynboetal2014,heintzetal2017}. During the prompt phase, the X-ray light curve exhibits three of prominent flares, of which the last one was also likely detected in the UVOT $u$ filter. In all UVOT filters, the decaying light curve is well-fit by a power law with the same index of $\alpha=0.9$, which is then flattened to $\alpha=0.54$ in the $g',r',i'$ and $z'$ filters between 0.33~and 3.5 days. After this, the lower data sampling does not allow for robust predictions, but a break very likely occurred between 3 and 20 days after $T_0$ before reaching the possible SN bump. Interestingly, Kann~et~al. found that bump occurrence could be associated with a color change towards blue, which is well-fit by the emergence of a thermal component. They proposed that this phenomenon could be attributed to the contribution of an associated SN, as previously put forth by Heintz~et~al. In this context, the putative SN must be considerably more luminous than the standard GRB~SNe/hypernovae and more akin to SN~2011kl; this could indicate that such luminous hypernovae may also accompany more standard LGRBs. However, in the case of GRB 140506A, there is no spectroscopic confirmation of a bright SN\endnote{A coeval spectrum at $T_0+52$ days was actually observed by Fynbo~et~al.~\cite{fynboetal2014}, but it is limited to $\sim$$660$ nm on the blue side. This missing piece of information might have revealed the contribution of the SN if present.}.}
\subsection{Other Possible Non-Standard GRB--SN Associations}
In some cases, data paucity or emission peculiarities of some GRBs make it difficult to asses whether or not an SN was actually accompanying a GRB and/or to characterize it in a consistent astrophysical framework. Nevertheless, these objects constitute an unresolved challenge that future studies will likely elucidate.
\subsubsection{GRB 210704A}
\textls[-30]{GRB 210704A~\cite{becerraetal2023} was an intermediate-duration GRB with $\tn\approx1$--$4$ s depending on the instrumental setup\endnote{Different instruments are sensitive to different energy bands and might not have detected every phase of the emission; see~\cite{becerraetal2023} for further details.}. In addition, the identification of the host galaxy is not unambiguous: an optical spectrum of GRB 210704A was observed at $T_0+1.1$ d, exhibiting a broad absorption, which, if interpreted as Ly$\alpha$, would correspond to a redshift of $z=2.34$. While the signal-to-noise ratio in this region is relatively low and does not allow for clear identification, the redshift inferred via the putative Ly$\alpha$ is in good agreement with that of the most probable host galaxy, given the position of the SN. Assuming this value for the redshift, the SN bump would correspond to an afterglow-corrected absolute magnitude at a peak of $M\approx-23.2$ mag in the UV band, which is even brighter than most SLSNe and is reached by some TDEs (see also~\cite{dongetal2016,holoeinetal2016,leloudasetal2016}). However, TDEs evolve on much longer time scales than GRB 210704A, and other possibilities cannot be ruled out.}
\subsubsection{GRB 221009A}
\begin{figure}[H]
    
    \includegraphics[width=0.7\textwidth]{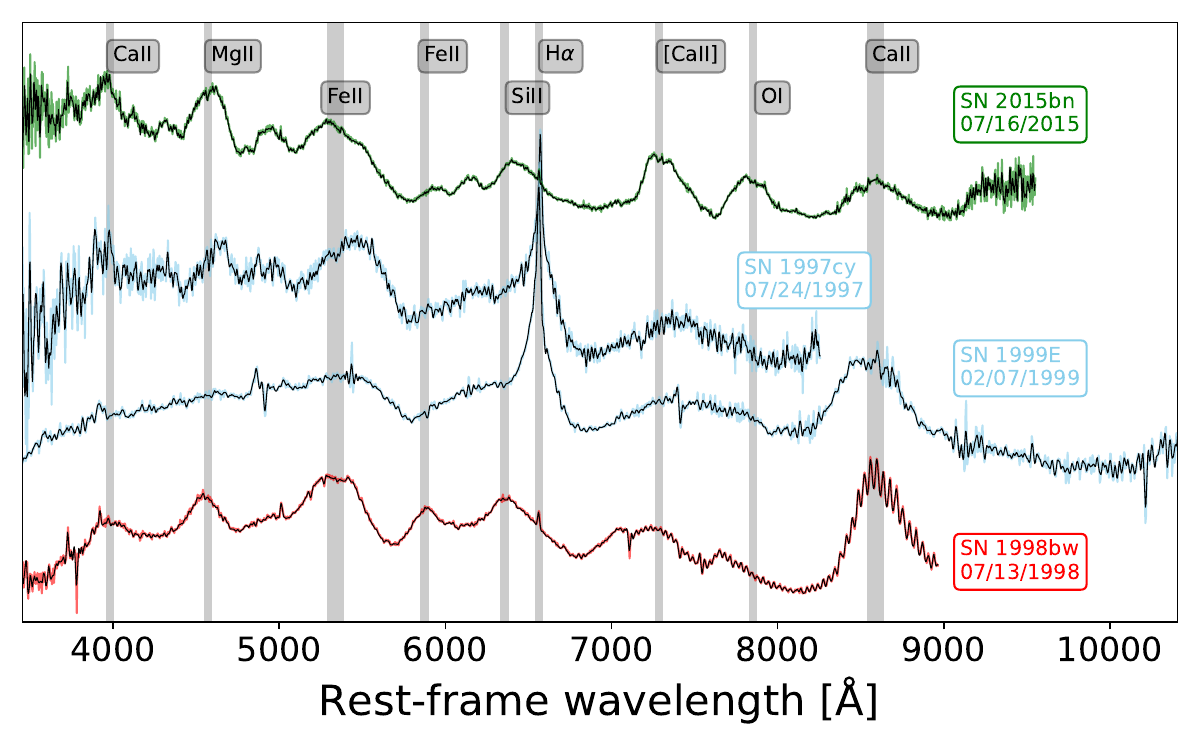}
    \caption{Spectral
 comparison of SN~1997cy and SN~1999E (light-blue spectra) with SN~1998bw (green spectrum) and SN~2015bn (red spectrum) (see the text for discussion). Each spectrum was smoothed with a Savitzky--Golay filter, and the smoothed spectrum is overplotted on the original in black and labeled with the SN name and the date of the observation. Some line identifications are marked with gray vertical lines and gray labels.}
    \label{fig:spectral_comparison}
\end{figure}

GRB 221009A is the brightest GRB ever observed, at a redshift of $z=0.151$. Due to its bright afterglow component and high dust extinction due to our galaxy, the presence of an associated SN in optical measurements could not be firmly assessed. However, late-time observations by the James Webb Space Telescope show that the spectrum at $\lambda<1.5\,\mu$m clearly deviates from a power law and provides evidence of several broad SN-like emission features (see Figure~\ref{fig:spectral_comparison}). 
The estimated $^{56}\mathrm{Ni}$ mass is $\sim$$0.09$ M$_{\odot}$, similar to GRB--SN prototype SN 1998bw at similar phases, suggesting that the SN of this exceptionally bright event was a typical GRB SN~\cite{Blanchard24}.
\subsubsection{LGRBs Mimicking SGRBs}
\label{sec:mimicking}
As anticipated, whether the $\tn$ alone is a good way to discriminate between short and long GRBs is matter of debate (see Section~\ref{sec:duration}). Here, we report on cases of a few GRBs that were classified as SGRBs but nevertheless show features more typical of LGRBs. This is the case of GRB~040924~\cite{donaghyetal2006,soderbergetal2006,wiersemaetal2008} and GRB~090426~\cite{antonellietal2009,nicuesaguelbenzuetal2011,thoeneetal2011a}. Depending on the energy band, GRB~040924 has a slow/intermediate $\tn\approx1$--$2$ s (depending on the energy band), and based on its duration and spectral characteristics, it can be classified as an SGRB~\cite{donaghyetal2006}. Therefore, it was surprising to observe a $\sim$$1$-mag wide SN bump in its afterglow light curve corresponding to the faintest GRB~SN known at that time as being $\sim$$1.5$-mag fainter than SN~1998bw~\cite{soderbergetal2006}. Furthermore, GRB~090426 had a soft high-energy spectrum and a bright afterglow, obeying an Amati relation typical of LGRBs. Hence, despite its duration of $\tn\lesssim2\,\mathrm{s}$, the evidence strongly supports a collapsar origin for these bursts. More recently, a similar object (GRB~200826A) was studied by Rossi~et~al.~\cite{rossietal2022}, supporting the idea that the GRB duration might not be enough to characterize its astrophysical scenario.

\subsubsection{GRBs Possibly Associated with SNe Interacting with CSM}
\label{sec:csmgrbsne}
In all cases examined thus far, the spectroscopically confirmed GRB~SNe are reminiscent of a stripped progenitor and have hydrogen-deficient spectra. In this section, we mention two possible exceptions, namely
SN~1997cy~\cite{germanyetal2000} and SN~1999E~\cite{rigonetal2003}, which have been associated with GRB~970514 and GRB~980910, respectively, and might further expand the variety of the observed GRB SNe.

GRB~970514 was discovered on 14 May 1997 with the BATSE satellite on board the Compton Gamma Ray Observatory; its pulse was marked by a fast-rising and exponentially decaying single peak~\cite{germanyetal2000}
. Due to its observed frame duration of $\sim$$0.2$ s, it was classified as an SGRB\endnote{However, Germany~et~al.~\cite{germanyetal2000} did not provide a clear definition of ``duration'' for GRB~970514.}. Nevertheless, GRB~970514 was included in the work of Woosley, Eastman and Schmidt~\cite{Woosley_1999} during a search for correlations between GRBs and unusually luminous SNe\endnote{This requirement stemmed from the high luminosity of SN~1998bw of $M_{\rm V}=-19.35$~\cite{galamaetal1998}.} (see also Section~\ref{sec:mimicking} for unusual associations of short-duration GRBs with SNe). GRB~970514 was the only one found to be in likely spatial and temporal association with the luminous SN~1997cy, whose magnitude at peak luminosity was $M_{\rm V}<-20.1$ mag, assuming $A_{\rm V}=0.00$ mag~\cite{turattoetal2000} for the galactic extinction. SN~1997cy was discovered as part of the Mount Stromlo Abell Cluster Supernova Search~\cite{reissetal1998} on 16 July 1997 in a compact and faint galaxy of the Sersic 40/6 cluster at a redshift of $z=0.059$~\cite{greenandgodwinandpeach1990}. Given the 4-month lag between the last pre-explosion image and the discovery image, there is no clear indication about the explosion date of SN~1997cy, and its association with GRB~970403 remains doubtful, but Germany~et~al.~\cite{germanyetal2000} estimated the probability\endnote{SN~1997cy exploded between 12 March 1997 and 15 July 1997. In this period, the BATSE satellite detected two GRBs within 2$\sigma$ from the position of SN~1997cy: GRB 970514 at 0.23$\sigma$ (corresponding to 0\degree.88) and GRB 970403 at 1.92$\sigma$ (corresponding to 17\degree.3) from SN~1997cy. Using the formalism of Wang and Wheeler~\cite{wangandwheeler1998}, a spatial coincidence within 0.23$\sigma$ corresponds to a probability of chance coincidence of $\sim$$1.7\,\%$.} of chance coincidence to be $\sim$$1.7\%$. The first optical spectrum of SN~1997cy was observed on 24 July 1997 with the Danish 1.5 m telescope~\cite{benettiandpizzellaandwheatley1997}, while two further spectra were taken on 9 August 1997 and 26 June 1998 with the Mount Stromlo and Siding Spring Observatory 2.3 m telescopes, respectively~\cite{germanyetal2000}. The characteristics of its spectrum are peculiar for a GRB SN/hypernova, as, together with more typical broad emission features from iron-peak and $\alpha$ elements resembling those showed by SN~1998bw~\cite{patatetal2001}, it presents a prominent multicomponent H$\alpha$ feature similar to SNe IIn (see Section~\ref{sec:snzoo}), bearing witness to the interaction of the SN ejecta with CSM\endnote{In the case of SN~1997cy, the intensities of these narrow/multicomponent lines change over time, thereby rendering them attributable to the SN itself (similar to the case of SN~1988Z~\cite{turattoetal1993}); in other cases (like in the spectra of SLSNe and LGRBs), narrow emission lines from these ions are also usually seen but with very different profiles and ascribed to the emission of the host galaxy.} (see Figure~\ref{fig:spectral_comparison}). Furthermore, the light curve of SN~1997cy shares some similarities in evolutionary time scales with that of the type IIn SN~1988Z~\cite{turattoetal1993}.

Although neither Germany~et~al. nor further studies could conclusively establish the link between GRB~970514 and SN~1997cy, it is worth noting that the case of SN~1997cy may not have been unique. SN~1999E~\cite{cappellaroetal1999,filippenkoetal1999,filippenko2000,rigonetal2003} was discovered by Roberto Antezana on 15 January 1999~\cite{perezetal1999}. Its spectro-photometric follow-up campaign was performed via several facilities in the optical and NIR at ESO La Silla, La Palma, and the Tonanzintla and Cananea observatories in Mexico. Similar to SN~1997cy, it was classified as an SN IIn due to its multi-component Balmer features in the spectra (see Figure~\ref{fig:spectral_comparison}); in particular, from the H$\alpha$, a redshift of $z=0.0261$ was measured, consistent with that inferred from the spectrum of the host galaxy. The light curve of SN~1999E resembles that of SN~1997cy across its evolution, although the latter evolves on quite slower time scales (see Figure 2 in~\cite{rigonetal2003}) and with an absolute magnitude at peak of $M_{V}<-19.5$ mag. As anticipated earlier, SN~1999E shares another peculiarity with SN~1997cy, as it was tentatively associated with GRB 980910, which was detected at 0.73$\sigma$ from the position of SN~1999E and translates to a probability of chance coincidence of 9.6\%. 

To the the best of the authors' knowledge, at the time of writing, there is no other GRB SN that has a clear spectral signature of CSM interaction. However, there are a few cases of very energetic and/or luminous SNe bearing imprints of interaction of the SN ejecta with CSM. For instance, GRB~011121 was discovered on 21 November 2001~\cite{greineretal2003,piroetal2001,wyrzykowskietal2001,olsenetal2001,brownetal2001,garnavichetal2003} at a redshift of $z=0.362$~\cite{infanteetal2001}, and it was classified as an LGRB due to its duration of $\tn=28$ s. \mbox{Garnavich~et~al.~\cite{garnavichetal2003}} found that a power law with an index of $\alpha=1.72$ thoroughly fits the $UBVRI$
afterglow data up to about three days after the burst, while the SED is well-fit by a power law with an index of $\beta=0.66$. Garnavich~et~al. found that these indices can be more easily interpreted with a scenario in which a jet shocks a uniform-density medium rather than an isotropic shock and, in particular, that they would be described by a wind-fed CSM, as also posited by Price~et~al.~\cite{priceetal2002}. In later phases, Bloom~et~al.~\cite{bloometal2002} showed that the light curves of the optical afterglow substantially deviate from the power-law behavior and are well described by an emerging bright SN, thereafter named SN~2001ke, peaking about 12~rest-frame days after the GRB with an absolute magnitude of $M_V\simeq-19.2$. Although this value is comparable with the absolute magnitude of SN~1998bw, the colors of this GRB~SN are bluer than usual hypernovae and more similar to those of SNe IIn, suggesting that CSM interaction played a role in powering it. In fact, the color evolution and the SED of SN~2001ke are remarkably similar to those of the strongly interacting type-IIn SN~1998S~\cite{filippenko1998}, although the latter on much slower time scales, as typical of hydrogen-rich SNe~\cite{garnavichetal2003}. Another possible example of a CSM-interacting hypernova is SN~2002ic~\cite{woodvaseyandalderingandnugent2002,hamuyetal2003}, which was classified as an SN Ia on the basis of the spectral features attributed to \ion{S}{{ii}} and \ion{Si}{{ii}}, typical of the optical spectra of SNe Ia around maximum luminosity. However, the late spectra of SN~2002ic were almost identical to those of SN~1997cy and SN~1999E~\cite{nomotoetal2004}. Benetti~ et~al.~\cite{benettietal2006} re-analyzed the classification of SN~2002ic as an SN Ia in more detail and found some inconsistencies in the absence of the H\&K \ion{Ca}{ii} lines and in the strength of the \ion{Si}{ii}+\ion{Fe}{ii} feature in the optical spectra of SN~2002ic. On this basis, they suggested that SN~2002ic was actually an SN Ic\endnote{In early epochs, the spectra of SNe Ia typically bear resemblance to those of SNe Ic (see e.g., Figure~1 in~\cite{hooketal2005} and~\cite{taubenbergeretal2006}).}. In addition, the light curve of SN~2002ic has a very bright peak absolute magnitude ($M_V<-20$ mag) for an SN Ia and presents a rebrightening, which Hamuy~et~al.~\cite{hamuyetal2003} attributed to the ejecta--CSM interaction. While these signatures would have favored the single-degenerate scenario for type-Ia SNe, they support the type-Ic classification for SN~2002ic, since more massive stars are expected to undergo significant mass loss prior to their explosion~\cite{benettietal2006}. While these associations are doubtful, it is intriguing to posit that luminous hypernovae may manifest as SNe IIn. This would not be implausible from an astrophysical point of view, since WR stars are known to eject CSM during the last stages of their evolution~\cite{pastorelloetal2008} (see also Shenar~\cite{shenar2024} for a recent review), and the presence of material around a GRB progenitor is a natural consequence of the core-collapse scenario.
\section{Environments}
\label{sec:environments}
\subsection{Progenitors}
Unraveling the identity of the progenitors of LGRBs, SNe Ic BL and SLSNe I is a complex problem, as it is a function of a huge number of unknowns with more than one possible solution. The strongest observational evidence supporting a common origin lies in the association of LGRBs with SNe Ic BL. As the latter are usually hydrogen- and helium-devoid, it is reasonable to deduce that the corresponding progenitors' layers were (almost) completely stripped off. 

{Two possible systems} may be considered to produce similar spectra after their explosions: (i) a single massive star experiences strong mass loss, evolving into a WR star or (ii) a progenitor star in a binary system undergoes mass transfer to a companion, losing its hydrogen and helium envelope before core collapse.  

{A viable progenitor scenario} should also be able to produce an LGRB, whose explosion mechanism usually requires (i) a rotating massive star collapsing into a compact object, typically a BH or a magnetar, and (ii) efficient energy extraction via stellar rotation to power the relativistic jet (see Sections~\ref{sec:grbs} and \ref{sec:magnetargrb}). Any of these mechanisms predicts that the progenitor star should retain a reasonably high angular momentum while simultaneously losing its H envelope and (at least a big fraction of the) He-envelope before the collapse. Therefore, WR stars seem to be a quite natural solution for this problem [see, e.g.,~\cite{conti1975,paczynski1967}]. However, not all WR stars meet these criteria, making the connection between WR stars and LGRBs non-trivial~\cite{Detmersetal2008}.

{Considering} the two possible scenarios for the progenitors of LGRBs associated with SNe Ic BL, in the single-star scenario, the progenitor is initially a very massive star ($\gtrsim$$25\, \mathrm{M}_\odot$) that undergoes significant radiation-driven stellar winds and/or rotational mixing during its lifetime. If the metallicity is high ($\gtrsim$$0.3 Z_\odot)$, strong line-driven winds efficiently strip the hydrogen and helium layers, leading to the formation of a classical WR star~\cite{woosley1993, Crowther2007}. However, at lower metallicities, mass-loss rates are significantly reduced, making it more difficult for a single star to shed its outer layers while retaining sufficient angular momentum to power an LGRB. In these environments, an alternative mechanism (quasi-chemically homogeneous evolution) has been proposed as a viable pathway~\cite{yoonandlanger2005, woosleyandheger2006}. In this scenario, rapid rotation drives strong internal mixing, preventing the formation of composition gradients and allowing the star to evolve directly into a compact, hydrogen-poor state. This pathway is strongly metallicity-dependent, since efficient quasi-chemically homogeneous evolution requires minimal angular momentum loss via stellar winds~\cite{yoonandlangerandnorman2006}. While the single-star pathway can produce WR progenitors with sufficient rotation to launch a GRB, it requires fine-tuned initial conditions. 

At high metallicities, excessive mass loss significantly removes a considerable amount of angular momentum and prevents GRB formation, while at extremely low metallicities, stars may fail to lose their envelopes entirely, leading instead to direct collapse into BHs with no explosion~\cite{hegeretal2003, MaederandMeynet2004}. As a result, alternative scenarios---particularly those involving binary evolution---are often invoked to explain the observed LGRB--SN Ic-BL connection. In the binary scenario, the progenitor is part of a close binary system where mass transfer via Roche-lobe overflow or common-envelope evolution strips the hydrogen and helium layers~\cite{Podsiadlowski2012,EldridgeandIzzardandTout2008}. This process allows even lower-mass stars ($\sim$$ 10$--$25\, \mathrm{M}_\odot$) to evolve into compact, helium-poor remnants resembling WR stars, even in environments where single-star mass loss would be inefficient~\cite{YoonandWoosleyandLanger2010, Sravanetal2020}. Importantly, binary interactions can also enhance rotation, either via tidal synchronization before mass transfer or by gaining angular momentum as material is transferred from the companion~\cite{cantielloetal2007, deMinketal2013}. One key advantage of the binary channel is the reduction in the metallicity dependence of the progenitor pathway, meaning that LGRBs can still form in moderately metal-rich environments where single-star pathways {{could be}} inefficient {{(see also Section}~\ref{sec:magnetars})}. Additionally, recent studies suggest that the majority of massive stars are born in binary or higher-order multiple systems~\cite{Sanaetal2012, MoeandDiStefano2017}, making binary-driven progenitors statistically more probable. Observationally, many LGRB host galaxies exhibit high specific star formation rates and low metallicities~\cite{Japeljetal2018, perleyetal2016}, conditions that could favor binary evolution due to their shorter lifetimes and increased likelihood of close interactions.

{In the context of quasi-chemically homogeneous evolution, Aguilera-Dena~ et~al}.~\cite{aguileradenaetal2018} {has used theoretical modeling to explore WR star progenitors,} which were also put forward by Nicholl~et~al.~\cite{nichollandguillochonandberger2017} as possible progenitors for SLSNe I. Aguilera-Dena~ et~al. computed detailed simulation of 5--100 $\msun$ quasi-chemically evolving progenitors. Depending on the mixing efficiency, they found that the analyzed WR stars may retain a considerable fraction of their core angular momentum during core collapse, as well as part of their original He-rich envelope. Lower-mass WR stars, typically in the range of $5$--$40\, \mathrm{M_\odot}$, have an angular momentum compatible with that expected for magnetar-driven SLSNe-I. In contrast, more massive WR stars, particularly those in the range of $40$--$100\, \mathrm{M_\odot}$, could plausibly give rise to LGRBs by forming BHs with sufficient angular momentum to launch relativistic jets. Nevertheless, higher-mass models may experience rotation-driven mass loss, for instance, through PPI phenomena, which could also explain the bumps seen in the light curves of many SLSNe I (see Section~\ref{sec:slsne}). Signatures of an interaction with a C/O-rich shell material would be then expected in some bumpy SLSNe I, of which a possible case could be represented by SLSN I SN~2018hti~\cite{linetal2020,fioreetal2022} displaying a ``boxy'' spectral feature~\cite{weiler2003,jerkstrand2017} in the optical (\ion{C}{ii} $\lambda\,6580$) and NIR (\ion{C}{ii} $\lambda\,9234$) spectra (see also Figure~14 in~\cite{fioreetal2022}).  Moreover, these features have been previously identified in some SLSNe-I, suggesting that at least a fraction of these events involve mass ejections from their progenitors before collapse~\cite{weiler2003,jerkstrand2017}. 

{{The properties of the canonical LGRB hosts were sometimes connected to an intrinsically young progenitor population that simply explodes earlier than other types of SNe}~\cite{leloudasetal2015, thoeneetal2015}. {However, Taggart and Perley}~\cite{taggartandperley2021} {suggested that these host properties could instead reflect an intrinsic preference for starbursting environments that favor the production of SLSNe I.} One possibility is that these transients arise from a top-heavy initial mass function that overproduces very massive stars, leading to a higher incidence of SLSNe~\cite{Dabringhausenetal2009}. An alternative explanation also discussed by~\citet{taggartandperley2021} is that the observed preference of SLSNe I for starburst galaxies could arise from collisional runaway processes in young and dense stellar clusters, as proposed by van den Heuvel and Portegies Zwart~\cite{vandenHeuvelandportegieszwart2013}. In this scenario, dynamical interactions in dense stellar environments could lead to the formation of extremely massive progenitors that are more likely to undergo PPI-driven mass loss and eventually explode as SLSNe I. These progenitor-driven mechanisms suggest that the formation of SLSNe I is inherently linked to the initial conditions of stellar evolution, particularly mass accretion histories, mass-loss mechanisms and rotational mixing effects that govern the late-stage evolution of massive stars. According to this view, the extreme mass loss necessary for some SLSN-I light-curve modulations may not arise solely from binary stripping or chemically homogeneous evolution but also from the complex interplay of stellar mergers, rotational mixing and PPI in progenitors exceeding $60$--$80\, \mathrm{M_\odot}$.

Progenitors of LGRBs and SLSNe I may then share similar characteristics, but the same phenomenon is rarely able to cause both. This reflects intrinsic differences of the angular momentum and of the magnetic field needed to power them ({see also Figures~2,3 in Kumar et al.}~\cite{kumaretal2024}). According to this view, the case of the SN~2011kl--GRB~111209A association (see Section~\ref{sec:11kl}) can be seen as a very unlikely case in which a progenitor star with intermediate features, similar to the classical GRB~SNe case, was able to power both. 

\subsection{Host Galaxies}
As CCSNe (in particular, SNe Ic) and LGRBs share a common astrophysical origin, i.e., the core collapse of massive stars, it is reasonable thinking that they also share similar host galaxies: however, it was shown by Fruchter~et~al.~\cite{fruchteretal2006} that CCSNe are usually found in environments more rich in metals than those hosting LGRBs. In fact, LGRBs are more often discovered in dwarf, low-metallicity host galaxies with a high specific star formation rate (sSFR, where $\mathrm{sSFR}\equiv\mathrm{SFR}/M_*$)~\cite{leflochetal2003,staneketal2006,modjazetal2008}. While LGRB hosts are usually poorer in metals than those of SNe Ic, SLSN I hosts are more similar to them and are usually dwarf, low-luminosity, low-metallicity galaxies---with a limiting metallicity of  \mbox{$Z\lesssim0.4$--$0.5\,Z_\odot$}~\cite{perleyetal2016,chenetal2017,schulzeetal2018}, above which SLSNe I are rarely observed and with a very high $\mathrm{sSFR}\approx10^{-9}\,\mathrm{yr^{-1}}$~\cite{stolletal2011,chenetal2013,lunnanetal2013,lunnanetal2014,japeljetal2016}. However, further studies pointed at an intrinsic difference between SLSNe I and LGRBs hosts. In particular, Leloudas~et~al.~\cite{leloudasetal2015} performed a detailed spectroscopic analysis of a sample of SLSN and LGRB host galaxies in the context of the SUSHIES\endnote{SUper-luminous Supernova Host galaxIES.} survey and showed that most galaxies hosting H-poor SLSNe ($\approx$$50\%$) were extreme emission-line galaxies (EELGs)
\endnote{Rather than a specific galaxy class, EELGs represent a rapid phase in the evolution of many galaxy types, like \ion{H}{ii} galaxies, blue compact dwarf galaxies, green pea galaxies, blueberry galaxies, and emission-line dot galaxies (see e.g.,~\cite{iglesiasparamoetal2022}) in the aftermath of a starburst.}~\cite{ateketal2011,amorinetal2014a,amorinetal2014b}, with even more ``extreme'' characteristics compared to LGRBs hosts. Leloudas~et~al. showed that the emission lines seen in SLSN-I hosts/EELGs are stronger than those seen in the spectra of CCSNe hosts and likely testify the presence of hard ionization fields, which, for SLSNe-I hosts/EELGs, are more intense than those of the host galaxies of GRBs (see Figure~2 of~\cite{leloudasetal2015}). These radiation fields, which are necessary to boost, e.g., \ion{He}{ii}, [\ion{Ar}{iv}] and [\ion{Fe}{iii}] lines~\cite{leloudasetal2015}, are usually indicative of the presence of WR stars. Conversely, the metallicity seems less helpful in discriminating between SLSNe-I hosts/EELGs and LGRB hosts, both being metal-poorer than those hosting normal CCSNe and SLSNe II (see Figure~2 of~\cite{leloudasetal2015}) and having a lower metallicity limit than SLSNe I  (e.g.,~\cite{grahamandfruchter2013,kruehleretal2015,schulzeetal2018}).

Furthermore, Lunnan~et~al.~\cite{lunnanetal2015} presented Hubble Space Telescope rest-frame UV imaging of SLSNe I hosts, and, for each of them, they studied the morphology and compared the position of the SN with the distribution of the UV light emitted from the host. Among the galaxies of their sample, they found that SLSNe-I hosts are usually compact and metal-poor with a high star formation surface density of $\Sigma_{\rm SFR}\sim0.1\,\mathrm{M_\odot\,yr^{-1}\,kpc^{-2}}$; in addition, SLSN I hosts often have an irregular morphology and are rarely present in a grand design spiral galaxy\endnote{With at least one notable exception: PTF10uhf~\cite{perleyetal2016}.}. However, even if these characteristics are broadly shared with the galaxies harboring LGRBs, Lunnan~et~al. found that the locations of LGRBs are much more correlated to the UV-bright regions of their hosts compared to SLSNe I (see also~\cite{angusetal2016,blanchardandbergerandfong2016}). While larger samples are required to statistically distinguish the two populations and draw a firm conclusion about this, the different correlation of SLSNe I and LGRBs with the UV bright spots of their host galaxies can be interpreted as an age/mass difference of their progenitors, assuming that the UV bright regions are robust tracers of star formation. Different environments and formation channels were also suggested by Angus~et~al.~\cite{angusetal2016}. In this work, NIR and UV data were used as an HST sample of SLSNe and compared by the authors with a sample of CCSNe and LGRBs, including ground-based optical host photometry and SED modeling, which allowed for determination of (s)SFRs and metallicities. This analysis pointed out that SLSNe I hosts present clear differences with those of CCSNe and LGRBs, the former having significantly lower masses and SFRs and higher compactness (see also~\cite{lymanetal2017}). Therefore, despite the similarities in metallicity\endnote{Angus~et~al.~\cite{angusetal2016} argued that this similarity might be due to selection effects in SLSNe~I, usually targeted to orphan (i.e., with a non-visible host in the images) events.}, Angus~et~al. suggested that LGRBs and SLSNe I owe their origin to different environments. 

However, whether or not metallicity is enough to explain the properties of SLSN I and LRGB hosts, in particular, the role of sSFR is still a matter of debate. Th\"one~et~al.~\cite{thoeneetal2015} studied spectroscopy of the galaxy hosting SLSN I PTF12dam and modeled the star formation history to estimate the epoch of its stellar population, which, in the case of PTF12dam, is even younger ($\sim$$3\,\mathrm{Myr}$) than LGRBs~\cite{leloudasetal2015}. They also concluded that SLSNe I are likely the result of the explosion of the most massive progenitors, showing that, contrary to the expectation due to the strong \ion{He}{ii} lines, the typical WR features are not seen in the host spectra. While this is not the case of many LGRB hosts~\cite{hammeretal2006,hanetal2010}, such evidence might disfavor WR stars as the main driver of the high radiation fields in SLSN I hosts. SLSNe I are possibly the endpoints of the very first and massive stars produced in a starburst at different locations compared to LGRBs~\cite{hsuetal2024}. In addition, correlations of both metallicity and sSFR across star-forming environments complicate 
the link between hosts and progenitors~\cite{tremontietal2004,salimetal2007}. Taggart and Perley~\cite{taggartandperley2021} attempted to solve the problem by comparing the properties of a sample of host galaxies of CCSNe taken from the All-Sky Automated Survey for Supernovae~\cite{shappeeetal2014} with those SLSNe and LGRBs and compared them against analogous data of LGBR and SLSN I hosts from the literature up to a redshift of 0.3, but based on their data, they were not able to break the degeneracy. Solving this degeneracy for LGRBs and SLSNe I will require more extended samples, possibly selected via untargeted galaxy~searches\endnote{Interestingly, Cleland~et~al.~\cite{clelandetal2023} considered a sample of galaxies from the SDSS with spectroscopic measurements to estimate the local galaxy density and found that SLSNe I usually explode in low-density environments. When applied to a set of simulated galaxies from the IllustrisTNG simulation, they found that densities suitable to reproduce SLSNe I hosts are better reproduced by constraining the host metallicity, while high sSFRs leave room to SLSNe I in high-density environments. Hence, they concluded that metallicity breaks the degeneracy in the case of~SLSNe I.}.
\section{Discussion and Conclusions}
\label{sec:discussion}
The link between GRBs and SNe has been robustly established through events like GRB~980425/SN~1998bw and GRB~030329/SN~2003dh, which especially provided compelling evidence for the association of LGRBs with SNe Ic BL, which is now well established with additional observations, such as GRB~060218/SN~2006aj~\cite{campanaetal2006, modjazetal2006, pianetal2006} and 
GRB~100316D/SN~2010bh~\cite{chornocketal2010,starlingetal2011}. SNe Ic BL present broad spectral features (see Figure~\ref{fig:spectral_comparison}), implying high photospheric velocity ($\sim$15,000--30,000 km s$^{-1}$~\cite{modjazetal2016}) and kinetic energy ($\sim$$10^{52}$ erg~\cite{iwamotoetal1998}) (see Section~\ref{sec:snzoo}). Both SNe Ic BL and LGRBs require a powerful central engine that can explain the extreme kinetic energy involved. 

Two primary models have been proposed for the central energy source: the collapsar (or BH-driven) scenario (see Section~\ref{sec:bhdriven}) and the magnetar scenario (see Section~\ref{sec:magdriven}). Both scenarios provide distinct mechanisms able to explain the observed phenomena, but they also highlight significant gaps in our understanding of progenitor conditions and explosion dynamics of GRB SNe and introduce new questions. The collapsar model attributes the origin of GRBs to relativistic jets launched from a BH--accretion disk system formed after the gravitational collapse of a massive, rapidly rotating star. This model successfully explains the energetics and observed durations of GRBs, but challenges remain in explaining the diversity of GRB--SN associations and the detailed properties of the progenitor stars. On the other hand, the magnetar scenario involves the formation of a fast-rotating, highly magnetized NS (also known as ``protomagnetar'') whose spin-down energy can drive both an energetic explosion and a collimated outflow. Studies have shown that magnetic fields exceeding $10^{14}-10^{15}$ G and rotational energies on the order of $10^{52}$ erg can power GRBs and their associated hypernovae (e.g.,~\cite{usov1992}). However, in this scenario, there are still uncertainties about the origin and amplification of such strong magnetic fields in progenitor stars, as well as their alignment with rapid rotation (e.g.,~\cite{margalitetal2018}).

GRB-associated SNe may differ from the standard SNe Ic BL, and in this review, we recall some examples (see Section~\ref{sec:pecgrbsne}). Notably, only one GRB has been found to be associated with an SLSN so far: GRB~111209A, an ULGRB with the longest burst duration ever measured, associated with SN~2011kl. This event has been cited to propose the extension of the classical GRB--SN connection to include brighter SNe and longer-lived LGRBs. This hypothesis is further supported by observational and theoretical arguments such as: (i) the spectroscopic resemblance of SLSNe I after $\sim$ weeks with SNe Ic BL at their maximum luminosity (see Section~\ref{sec:slsne}), (ii) the similarities of their host-galaxy environments (see Section~\ref{sec:environments}) and (iii) the compatibility of a magnetar scenario for both of them (see Section~\ref{sec:magnetarsn}). The last point (iii)  merits further discussion. In fact, while a magnetar with a magnetic field $>$$10^{15}\,\mathrm{G}$ is, in principle, capable of boosting the duration of a GRB and conveying the energy to power a highly collimated (but weakly relativistic) jet~\cite{komissarovandbarkov2007,bucciantinietal2007,bucciantinietal2008}, this same object would hardly power a bright SN, which usually requires a polar magnetic field of $B\approx0.01$--$1\times10^{15}\,\mathrm{G}$~\cite{kasenandbildsten2010,prasannaetal2023} (see also Equation~(\ref{eq:sdlum3})\endnote{In principle, even a weaker magnetic field of $<$$10^{13}\,\mathrm{G}$ should allow for an even brighter SNe, which is usually not observed. This could also be due to the amplification mechanism of the magnetic field (like dynamo processes~\cite{thompsonandduncan1993}), allowing for a given range of magnetic-field strengths~\cite{metzgeretal2015}.}). Moreover, irrespective of the assumed model, while LGRBs typically need to release energy on a time scale of $\sim$$10^2$--$10^3$ s to launch an ultrarelativistic jet, in the case of a core-collapse (SL)SN, the engine should fuel the {ejecta} $\sim${weeks} after the core collapse (as discussed later in this section).

Metzger~et~al.~\cite{metzgeretal2015} investigated the applicability of the magnetar paradigm to provide a consistent explanation for events like GRB 111209A/SN~2011kl, i.e., both for the ULGRB and the SLSN~I. The authors used the  spin-down luminosity at $t=0$ as a proxy for the jet luminosity, i.e., $L_j\approx L_{\rm sd}(t=0)$ (see Equations~(\ref{eq:sdlum1}) and (\ref{eq:sdlum2})), which translates to an observed isotropic luminosity of $L_{\rm iso}=\epsilon_\gamma f^{-1}_{\rm b}L_{\rm sd}$, where $\epsilon_\gamma$ and $f_{\rm b}$ are the conversion efficiency and the beaming factor, respectively. Furthermore, synthetic SN light curves are computed following a diffusion scheme similar to~\cite{arnett1982}, assuming the sum of $^{56}$Ni decay and $L_{\rm sd}$ as the energy source powering the SN luminosity. In Figure~1 of~\cite{metzgeretal2015}, luminosities and time scales computed with this scheme (see~\cite{metzgeretal2015} for further details) are plotted as contours in the $B,P$ plane and compared with some estimates obtained based on observational GRB-SNe/SLSNe-I data. In particular, in the top panel of this figure, some regions are marked with boxes and arrows that define the loci where normal LGRBs (top-left corner, low $P$, high $B$), ULGRBs ($t_{\rm sd}\gtrsim10^3$--$10^4\,\mathrm{s}$) and $^{56}$Ni decay-powered SNe (either $B\approx10^{15}\,\mathrm{G}$ or very low $B$ and large $P$) exist. For the latter, the contribution of the magnetar is subdominant, while for $10^{12}\,\mathrm{G}\lesssim B\lesssim10^{15}\,\mathrm{G}$, it becomes the major power source of the SN and reaches typical SLSN luminosities. Interestingly, the case of GRB~111209A/SN~2011kl (indicated by a red dot in Figure~1 of~\cite{metzgeretal2015})
is found at the intersection between the ULGRB and the SLSN~I regions. Metzger~ et~al. predicted $B=3\times10^{14}\,\mathrm{G}$ and $P=2\,\mathrm{ms}$ for GRB~111209A/SN~2011kl, fixing $M_{\rm ejecta}=3\,\mathrm{M_\odot}$, which is about one-third of the average ejecta mass of the SLSNe~I of $\sim$$10\,\mathrm{M_\odot}$ measured by Nicholl~et~al.~\cite{nicholletal2015}. While $M_{\rm ejecta}=3\,\mathrm{M_\odot}$ is a mere assumption in the model of Metzger~et~al., it is consistent with other estimates of the ejecta mass of SN~2011kl~\cite{greineretal2015,mazzalietal2016}. According to this view, a sub-average ejecta mass for GRB~111209A/SN~2011kl might have allowed the jet to break out. This could also explain why a burst is not observed in many other SLSNe I, which are commonly associated with higher-mass ejecta. 

Metzger~et~al.~\cite{metzgeretal2015} showed how the magnetar picture also applies to classical GRB SNe made by LGRBs and SNe Ic BL, the latter usually being described with $^{56}$Ni decay. Such a hypothesis was previously proposed by Mazzali~et~al.~\cite{mazzalietal2014,mazzalietal2016} and Wang~et~al.~\cite{wangetal2017b}. In particular, Mazzali~et~al.~\cite{mazzalietal2014} suggested that classical GRB SNe could also be reasonably explained by millisecond magnetars, as their kinetic energies are typically limited by the theoretical maximum rotational energy of a magnetar, which can be estimated with Equation~(\ref{eq:rotenergy}). Moreover, Mazzali~et~al.~\cite{mazzalietal2016} analyzed the spectral properties of SN~2011kl in the context of SLSNe~I spectra and discussed the following results: The only spectrum available for SN~2011kl (see Section~\ref{sec:11kl}) has a cooler continuum than that of other SLSNe I (in this case, iPF13ajg) and expansion velocities ($\sim$20,000$\,\mathrm{km\,s^{-1}}$) more similar to those of classical GRB SNe. Despite its lower continuum temperature, the ejecta of SN~2011kl were likely energized by a non-thermal mechanism due to the presence of W-shaped absorptions.  
Such absorption features, if correctly identified as transitions from \ion{O}{ii} (see Section~\ref{sec:slsne}), have a remarkable excitation potential of $\sim$$25\,\mathrm{eV}$ and likely require a non-thermal excitation mechanism. Therefore, the magnetar scenario seems to be an intriguing possibility to account for the spectroscopic features of SN~2011kl if one assumes that the magnetar-driven shock (see Section~\ref{sec:magnetarsn}) may inject energy in the SN ejecta\endnote{In principle, a magnetar-driven shock injecting energy into the ejecta could also synthesize more $^{56}$Ni, but in the case of SN~2011kl, the reduced UV suppression~\cite{greineretal2015} makes the $^{56}$Ni-driven scenario more challenging.}.
More recently, non-relativistic 2D MHD simulations of rapidly rotating protomagnetars by Prasanna~et~al.~\cite{prasannaetal2023} supported the idea that a magnetar with spin periods of $4\,\mathrm{ms}$ and a magnetic field $\gtrsim$$10^{15}\,\mathrm{G}$ can release up to $5\times10^{51}\,\mathrm{erg}$ of energy\endnote{Higher values of the total energy released by the PNS ($<$$10^{52}\,\mathrm{erg}$) can be achieved for lower PNS spin periods and higher polar magnetic fields.}, making it a promising energy source for GRBs. 

Metzger~et~al.~\cite{metzgeretal2015} showed how magnetar could represent a common framework to interpret the connection of LGRBs and even ULGRBs with SNe Ic BL and SLSNe~I, trying to answer the question of whether a single engine can power both events, but they did not investigate the possibility of the two occurring in the same event
. A significant step forward in examining such a hypothesis was made by Margalit~et~al.~\cite{margalitetal2018}, who proposed a unified model in which a single central engine---in this case, a magnetar---can simultaneously power both a GRB jet and an SLSN~I. In their model, the misalignment between the rotation ($\Omega$) and magnetic dipole ($\mu$) axes is a key factor in providing a mechanism for thermalization of the spin-down power through reconnection in the striped equatorial wind~\cite{Lyubarsky2003}. This component can be the source of the thermal emission that powers the SN, while the remaining unthermalized energy launches the relativistic jets.  Before this work, the problem of simultaneously launching a collimated relativistic jet and an isotropic thermal SN had already been discussed by some authors (see, for instance,~\cite{thompsonandchangandquataert2004}). However, none of them investigated a possible mechanism for the magnetar energy partition. To address this question, Margalit~et~al. first assumed that all of the magnetar rotational energy goes into electromagnetic spin-down, neglecting gravitational wave emission (see Section~\ref{sec:magnetarsn} and Equation~(\ref{eq:spindown0})), then employed a semi-analytical model to partition the spin-down luminosity into both thermal and magnetically dominated components depending on the misalignment angle ($\alpha$) between the magnetar rotation and magnetic axes. The magnetic ($f_\mathrm{j}$) and thermalized ($f_\mathrm{th}$) energy fraction can be respectively approximated as follows:
\begin{align}\label{eq:misalignment_1}
f_\mathrm{j} & = \frac{3}{2} \int_0^{\pi/2}\chi(\theta;\alpha)\sin^3{\theta}\,\mathrm{d}\theta \, , \\
\label{eq:misalignment_2}
    f_\mathrm{th}&\simeq \frac{1.025\alpha}{(0.636+\alpha^4)^{1/4}}\, ,   
\end{align}
where $\chi(\theta;\alpha)$ is the fraction of wind power remaining in Poynting flux at latitude $\theta$, with $\alpha$ and $\theta$ given in radians. Their model assumes that a tilted rotator will divide its spin-down power between a magnetic and a thermal component, where $f_\mathrm{th}$ determines how much is available to power the SN. Equation~(\ref{eq:misalignment_2}) shows that thermalization increases with greater $\alpha$ values. Thus, for small misalignment angles ($\alpha \approx 0$), the energy is mainly magnetic and powers the jet (GRB-dominated scenario), while for large misalignment angles ($\alpha \approx 90^\circ$), most of the energy is thermalized and contributes to the SLSN. The competition between $f_\mathrm{j}$ and $f_\mathrm{th}$ competition depending on $\alpha$ {\mbox{is shown in Figure~2 of}~\cite{margalitetal2018}}
.
Margalit~et~al. used a collimated-jet model~\cite{Brombergetal2011} to investigate jet propagation and breakout through an exploding stellar profile and expanding SN ejecta. One of the major issues still discussed with respect to the connection of SNe and LGRBs within one event is the duration necessary for the engine to reach the peak luminosity (see Equation~(\ref{eq:sdtime})), which is days or longer for SLSNe ($\sim$$10$~d~\cite{metzgeretal2015}), compared to minutes or even less for GRBs ($\sim$$100$~s). As both GRBs and SLSNe~I should be powered by the same central engine, the efficiency of energy partition in different components plays an important role in shaping the output, as well as the ability of the jet to break out of
 the stellar progenitor faster than the SN shock.

Margalit~et~al.~\cite{margalitetal2018} considered three breakout regimes depending on the dimensionless jet luminosity parameter ($\Tilde{L}_\star$)\endnote{For the analytical expression of $\Tilde{L}_\star$, refer to Equations~(15) and (16) in~\cite{margalitetal2018}.}: (i) the ``strong jet regime'' when the $\Tilde{L}_\star\gtrsim 1$; (ii) an intermediate regime when $\Tilde{L}_\star\sim1$, where the jet-escape time scale is comparable to the time necessary for the SN shock to reach the stellar surface; and (iii) the ``weak jet regime'' when the $\Tilde{L}_\star\lesssim1$. In the first case, the jet luminosity is large enough to break out before the SN blast wave has any significant effect on the outer layers of the star and is the one usually considered in the literature~\cite{Brombergetal2011}, as the jet can be treated as propagating within a hydrostatic stellar environment. The case when  $\Tilde{L}_\star\sim1$ marks the transition between the jet breakout and the expanding envelope. Researchers found that when the jet luminosity is below $L_\mathrm{j}\sim 3\times 10^{47}$~erg~s$^{-1}$, it cannot break out before the SN ejecta expands significantly. In the ``weak jet regime'', which is relevant for SLSNe, they identified two conditions for successful jet emergence: (i) the jet-head velocity must exceed the expanding SN ejecta velocity ($v_\mathrm{h}\gg v_\mathrm{ej}$~\cite{QuataertandKasen2012}), and (ii) the jet must be stable against MHD instabilities, particularly kink instability. They found that these conditions imply a minimum luminosity threshold for jet breakout ($L_\mathrm{j}\gtrsim 10^{46}$~erg~s$^{-1}$), suggesting that many SLSNe could be accompanied by hidden GRB-like jets. The condition on $L_\mathrm{j}$ can also be expressed in terms of the total energy of the engine ($E_\mathrm{e})$\endnote{$E_{\rm e}$) can be expressed in terms of the jet energy as $E_{\rm j}=f_{\rm j}E_{\rm e}/2$, where the factor of 2 accounts for a bipolar jet~\cite{margalitetal2018}.} as follows:
\begin{equation}\label{eq:P0_Margalit}
    f_\mathrm{j}E_\mathrm{e}\gtrsim 0.195E_\mathrm{SN}\, .
\end{equation} 
Assuming the magnetar scenario, $E_\mathrm{e}$ can be evaluated using Equations~(\ref{eq:rotenergy}) and  (\ref{eq:P0_Margalit}), leading to a maximum initial spin period of $P_0\simeq 10$~ms for a typical SN explosion energy of $E_\mathrm{SN}\sim10^{51}$~erg. We note that for a more canonical hypernova energy ($E_{\rm HN}=10^{52}\,\mathrm{erg}$) and a reasonable assumption on $f_{\rm j}=0.55$ (as in the case of LSQ14bdq; see later in the text), this value can be lowered to $\sim$$2.7\,\mathrm{ms}$. Below this maximum value\endnote{This condition also depends on the value of $f_{\rm j}$ and on the jet Lorentz factor; see Equation~(22) in~\cite{margalitetal2018} for further details.}, a magnetar can power both an SLSN and a relativistic jet.

Margalit~et~al. also estimated the jet breakout time, finding that if $E_\mathrm{e}>E_\mathrm{SN}$, the jet can emerge within hours after the explosion. This result shows that jets with low luminosity of the same order of magnitude as the engine luminosities needed to power SLSNe are able to escape the SN ejecta on time scales comparable to the engine lifetime. This is consistent with the case of GRB~111209A associated with SN~2011kl~\cite{greineretal2015}.
One of the most significant contributions of this work is the prediction of the observational signatures that could come from these off-axis jets that preclude the association of a GRB with the corresponding SN. This finding then explains why many SLSNe are detected without an accompanying GRB, supporting the connection between SLSNe and ``hidden'' GRBs. Researchers found that the breakout of a transrelativistic ``shocked-jet'' cocoon component can cause a shock heating of the surrounding material, leading to a short-lived UV flare that lasts a few hours and reaches a luminosity of $L_\mathrm{peak} \sim 10^{44}$--$10^{45}$~erg~s$^{-1}$. Because of the short duration, these events are short-lived and difficult to detect without high-cadence surveys. However, the authors suggested that wide-field UV survey missions such as ULTRASAT~\cite{Ganotetal2016} should be able to detect them. In their model, they also suggested a mechanism able to explain the early maxima observed in SLSNe light curves that cannot be accounted for by the breakout emission~\cite{Leloudasetal2012,nicholletal2015,smithetal2016}. Energy and momentum being effectively dissipated\endnote{Margalit~ et~al. did not investigate the properties of this mechanism, but several have been proposed (see e.g.,~\cite{Morsonyetal2010,barniolduranetal2017,lazzatietal2011}).} at the jet--ejecta interface can result in thermal wind rendering an optical/UV peak in the SLSNe light curves before the maximum luminosity of the SN. With their wind model, Margalit~et~al. were able to reproduce the deeply sampled $r$-band light curve of the double-peaked\endnote{LSQ14bdq is not the only SLSN I in which a pre-maximum bump shows up. In fact, this is also the case of SN~2006oz~\cite{Leloudasetal2012}, DES14X3taz~\cite{smithetal2016} and possibly of SN~2018hti~\cite{fioreetal2022}.} SLSN I LSQ14bdq~\cite{nicholletal2015b} (see Figure~4 in~\cite{margalitetal2018}). 
In addition, they predicted that off-axis jets could interact with the ambient medium and cause a late-time radio afterglow, similar to orphan GRB afterglows.  

In addition, detailed 3D general relativistic MHD simulations of proto-magnetar jets in a CCSN scenario conducted by Shankar~et~al.~\cite{shankaretal2021} further support the viability of the magnetar scenario for both GRBs and SLSNe. Their study focused on extracting jet properties---such as energy, duration and collimation---from a proto-magnetar and examining how these parameters influence both SN and GRB observables. They demonstrated that magnetar-driven jets can naturally provide the high kinetic energy (of the order of $\sim$$10^{52}$~erg) needed to explain SNe Ic, reinforcing the idea that such jets could serve as a unifying engine behind both GRBs and SLSNe. By employing hydrodynamic and radiation transfer simulations, they further showed that the light curves and spectra produced by their models are consistent with observed SNe-type Ic-BL, thereby strengthening the connection between relativistic jets and highly energetic SNe.

A key aspect of their findings concerns the jet collimation and breakout conditions, providing additional justification for the energy partitioning mechanism proposed by Margalit~et~al.~\cite{margalitetal2018}. While the model of Margalit~et~al.~\cite{margalitetal2018} theoretically predicts that the misalignment between the rotation and magnetic axes of a magnetar can naturally split its spin-down energy into collimated and isotropic components, Shankar~et~al. offered a numerical confirmation of this effect by explicitly showing how jet opening angles influence observability. In their simulations, only jets with a sufficiently small half-opening angle ($\theta_{\rm eng} \sim 11^\circ$) were able to produce GRBs, whereas wider jets ($\theta_{\rm eng} \sim 17^\circ$) resulted in SNe-type Ic-BL without an associated $\gamma$-ray signal. This supports the argument that many SNe could harbor relativistic jets that fail to break out or remain hidden due to unfavorable viewing angles. Moreover, the connection between jet half-opening angles and GRB detectability implies that relativistic jets could be present in a significant fraction of SLSNe, even when a $\gamma$-ray signal is absent. This naturally explains why GRBs are not observed in all SNe-type Ic-BL or SLSNe events, as a large fraction of these jets may either be too weak to emerge from the SN ejecta or be oriented away from the observer’s line of sight. Additionally, their models predict that these hidden GRB-like events could still leave observable imprints, such as late-time radio afterglows or shock-breakout UV signatures, which could be detected by high-cadence wide-field surveys. Thus, their work provides strong numerical evidence in favor of a unified magnetar-driven framework where the observable outcome---GRB, SN or both---depends primarily on the jet properties and viewing geometry. It is important to specify that Margalit~et~al.~\cite{margalitetal2018} focused on the connection between SLSN~I and LGRBs, while they did not explicitly suggest that SLSN~I can be connected with ULGRBs. However, they mentioned the case of ULGRB GRB~111209A, which was associated with a highly luminous and short-lived SN resembling an SLSN, as a possible direct link between GRBs and SLSNe~I. Overall, they provided a compelling unifying framework for the GRB--SLSN connection, demonstrating how they can be two different manifestations of a central engine-powered event. For their model, they used a magnetar as central engine and showed that a misalignment between the rotational and magnetic axes can naturally partition its spin-down energy between a collimated relativistic jet and an isotropic thermal component, allowing a single system to power both a GRB and an SLSN. Their results suggest that many SLSNe could host relativistic jets that remain undetected due to weak breakout conditions or off-axis orientation. Moreover, they also provided observational signatures that could help in identifying hidden GRB-like activity in these events. With their study, Margalit et al were able to provide a positive answer to the following question: {``Can a single system power both GRBs and SNe within one event'',} even specifically even when the GRB is not observed? However, they did not rule out the collapsar scenario, since their findings for the weak jet breakout and associated observational signatures could also be applied to a BH engine model, as long as a similar energy partition mechanism takes~place.

The observational evidence, along with detailed theoretical models and simulations, supports the magnetar scenario for the GRB--SN connection, but some challenges remain. The formation of a magnetar requires precise conditions, including a rapidly rotating core and strong differential rotation during collapse. Additionally, the transition from magnetar-driven jets to the observed GRB emission involves complex magnetic dissipation processes that are not fully understood~\cite{mostaetal2015}. Furthermore, the magnetar scenario struggles to account for the most rapidly variable GRB light curves, which may need the complex accretion dynamics of BH-driven jets. Alternatively, the PPI in the extended helium progenitor model proposed by Moriya~ et~al.~\cite{Moriyaetal2020} for GRB~111209A/SN~2011kl also ties together the extended progenitor structure, the long-duration GRB and the luminous SN in a self-consistent framework and cannot be ruled out.
However, this model also presents challenges, since it is not the only one able to provide a good modeling of GRB~111209A/SN~2011kl. The fallback accretion model, for instance, can provide prolonged energy injection, potentially accounting for ULGRBs without requiring an extended progenitor envelope~\cite{margalitetal2018}. Additionally, jet--cocoon interactions within the stellar envelope have been proposed as a mechanism for producing rapidly evolving luminous transients (as discussed later), independent of magnetar spin-down. These models naturally explain the diversity in observed GRB-SN light curves and the presence of relativistic outflows, even in cases where direct jet emergence is inhibited. Another difficulty that can arise for Moriya~et~al.'s model is that forming extended helium progenitors remains a challenge for the PPI mechanism. While Moriya~et~al.~\cite{Moriyaetal2020} suggested that such stars can retain their extended structure until collapse, uncertainties in mass loss due to strong WR winds and binary interactions raise concerns about their viability. Additionally, the expected CSM signatures from repeated pulsations are not always observed in hypernova spectra (see Section~\ref{sec:csmgrbsne} for possible exceptions), making it difficult to confirm whether PPI-driven mass ejections play a significant role in GRB SNe. Moreover, while PPI is more favorable in low-metallicity environments, some GRB SNe and SLSNe are found in higher-metallicity galaxies, suggesting that other processes may be responsible for the observed explosion characteristics.

BSGs have also been put forward as viable progenitors for events like GRB~111209A/ SN~2011kl. This hypothesis follows quite straightforwardly from the ULGRB perspective (see Section~\ref{sec:ulgrbs}) but needs a way to produce a very luminous SN like SN~2011kl. In fact, BSG explosions usually yield failed-SNe explosions, although some SNe events have been connected to a BSG explosion~\cite{Arnettetal1989}. Nakauchi~et~al.~\cite{nakauchietal2013} suggested that a luminous SN-like bump in the afterglow light curve of a BSG-driven ULGRB can be caused by the energy dissipated by the jet head in the hot cocoon (see Section~\ref{sec:phenomenology}), the so-called cocoon-fireball photospheric emission. Similar to the SN case, the cocoon is radiation-dominated and non-relativistic, resulting in a photospheric emission. However, due to the extended envelope of the progenitor, Nakauchi~et~al. predicted that the cocoon emission should look like SNe IIP (see Section~\ref{sec:snzoo}), as they included hydrogen-rich matter from the extended BSG envelope. However, a type-II SN appearance is in sharp contrast with the hydrogen-poor spectra of SN~2011kl (see Section~\ref{sec:11kl}). The connection between BSG explosion and eventually luminous SNe has been further investigated considering mechanisms other than the cocoon-fireball photospheric emission. We mention the model proposed by Fisher~et~al.~\cite{fisheretal2018}, who speculated that collapsing BSGs may allow for quark--deconfinement phase transition, which can then trigger successful and bright SN explosions. According to this scenario, the SN shock wave is rejuvenated by the release of latent heat of the hadron--quark phase transition, and the ejected, hydrogen-rich matter of the envelope can act as CSM and let a bright interacting SN shine. Assuming that such an explosion can also power a ULGRB\endnote{This was not investigated by Fisher~et~al.~\cite{fisheretal2018}.}, this scenario struggles to explain H-poor SLSNe like SN~2011kl, making the magnetar scenario more likely (see Section~\ref{sec:11kl} for the challenges of the magnetar scenario posed by this object).

These mechanisms offer possible pathways to produce very luminous SNe and possibly link them with GRBs; nevertheless, they also underscore the inherent difficulties in explaining the full range of observed properties in a single framework. Theoretical models attempting to unify ULGRBs and SLSNe~I often introduce complex dependencies on progenitor structure, explosion dynamics and energy deposition mechanisms, many of which require fine-tuned conditions that may not be naturally realized in most stellar collapse events. In particular, the apparent mismatch between predicted and observed spectroscopic features in SN~2011kl, as well as the uncertain role of jet-driven and magnetar-driven energy contributions, suggests that no single progenitor scenario can straightforwardly account for both components. This brings us to a broader issue: although very luminous hypernovae can be associated with LGRBs and ULGRBs, the theoretical scenarios analyzed in this review for powering both an SLSN~I and a GRB within the same event face significant theoretical challenges, i.e., we have shown that they require fine-tuned conditions, which are difficult to achieve in a single event. This last point tends to disfavor a `superluminous hypernova--GRB' connection similar to the well-established one with SNe Ic BL, and it might be supported by the paucity of SLSNe~I-GRB observations. Furthermore, the estimated rates of SLSNe~I ($91^{+76}_{-36}\,\mathrm{Gpc^{-3}\,yr^{-1}}$) and ULGRB ($\approx$$30\,\mathrm{Gpc^{-3}\,yr^{-1}}$) were shown to be comparable within $z<1$~\cite{prajsetal2017}; this similarity is reinforced if we restrict ourselves to the case of SLSNe~I within $z<0.89$ ($\approx$$40\,\mathrm{Gpc^{-3}\,yr^{-1}}$)~\cite{zhaoetal2021}. If only a fraction is powered by a magnetar, selecting those fulfilling a condition similar to (\ref{eq:P0_Margalit}) with a successful GRB directed towards us can significantly lower the combined rate of GRB superluminous hypernovae and explain why, after about 20 years of Swift observation, only one case has been observed. Future observations with advanced facilities like the Vera C. Rubin Observatory and the James Webb Space Telescope will provide detailed data on GRB SNe, enabling more precise tests of the magnetar model. Numerical simulations will also continue to refine our understanding of the interplay between magnetars, relativistic jets, and SNe.

\vspace{6pt} {}

\authorcontributions{{All the authors worked on the conceptualization and writing of this manuscript. All authors have read and agreed to the published version of the manuscript.} 
}
\funding{{This research received no external funding} 
}
\dataavailability{{Not applicable.} 
} 

\acknowledgments{We thank the anonymous referees for their insightful comments. {A. F. acknowledges funding by the European Union – NextGenerationEU RFF M4C2 1.1 PRIN 2022 project ``2022RJLWHN URKA'' and
by INAF 2023 Theory Grant ObFu 1.05.23.06.06 ``Understanding R-process \& Kilonovae Aspects (URKA)''}. A.F. and G.S. acknowledge the support of the State of Hesse within the ELEMENTS Research Cluster (Project ID 500/10.006). We thank Sylvio Klose, Ana Nicuesa Guelbenzu and Stefano Benetti for interesting discussions about this work. We also thank Massimo Turatto, Jochen Greiner and Andrew Levan for kindly allowing us to reproduce the figures used throughout the present work, and we thank Fuyuan Zhao for the permission granted to quote~\cite{zhaoetal2006}.}

\conflictsofinterest{The authors declare no conflicts of interest. The funders had no role in the design of the study; in the collection, analyses or interpretation of data; in the writing of the manuscript; or in the decision to publish the results.} 
\appendixtitles{no} 
\appendixstart
\appendix

\begin{adjustwidth}{-\extralength}{0cm}
\printendnotes[custom]
\end{adjustwidth}
\begin{adjustwidth}{-\extralength}{0cm}
\reftitle{References}

\PublishersNote{}
\end{adjustwidth}
\end{document}